\documentclass[
preprint,
showpacs,preprintnumbers,
nofootinbib,
 aps,
prd,
floatfix,
titlepage,
showkeys
]{revtex4-2}

\usepackage{graphicx} % Required for inserting images
\usepackage{color}
\usepackage{amsmath}
\usepackage{braket}
\usepackage{hyperref}
\usepackage{bbold}
\begin{document}
%\title{Full Edge Control of Bulk Energy Gap and Topological Phase in Armchair Graphene Nanoribbons}
\title{Full edge control of bulk energy gap in armchair graphene nanoribbons}
\author{Lin Wang}
\email{l.wang@fz-juelich.de}
\affiliation{
 Institute for Advanced Simulation (IAS-4)\\
 Forschungszentrum J\"ulich, Germany
}%
\author{Thomas Luu}
\email{t.luu@fz-juelich.de}
\affiliation{
 Institute for Advanced Simulation (IAS-4)\\
 Forschungszentrum J\"ulich, Germany
}%
\author{Ulf-G. Mei{\ss}ner}
\email{meissner@hiskp.uni-bonn.de}
\affiliation{Helmholtz Institut f\"ur Strahlen- und Kernphysik, Bethe Center for Theoretical Physics and Cluster of Excellence  ``Color meets Flavor'', 
Universit\"at Bonn, D-53115 Bonn, Germany}
\affiliation{ Institute for Advanced Simulation (IAS-4)\\
 Forschungszentrum J\"ulich, Germany}
%\affiliation{Peng Huanwu Collaborative Center for Research and Education, International Institute for Interdisciplinary and Frontiers, Beihang University, Beijing 100191, China}

\begin{abstract}
In solid-state physics, the ability to open and tune an electronic energy gap provides a direct route to controlling charge transport and accessing distinct electronic phases. In solids with physical boundaries, a bulk energy gap formed exclusively by extended bulk states is conventionally expected to be governed primarily by the interior bulk region, with only a minor contribution from the boundaries. Here we show that, in stark contrast to this conventional expectation, the bulk gap of armchair graphene nanoribbons in the $N_w=3p+2$ family with $p\in\mathbb{Z}\ge 0$ can be fully controlled through the hoppings \emph{solely at the boundaries}. This full boundary control originates from an exact cancellation between the gap contributions of long-range hoppings along and across the ribbon in the bulk region. It persists in the presence of Hubbard interactions and holds irrespective of ribbon width in this family. Our findings establish boundary dynamics as an unexpectedly powerful means of manipulating bulk electronic properties, enabling width-preserving engineering of one-dimensional (1D) band structures and topological phases, with potential applications in qubits, artificial quantum spin chains, topological quantum dots and other quantum devices.

%The ability to engineer topologically distinct materials opens the possibility of enabling novel phenomena in low-dimensional nano-systems, as well as manufacturing novel quantum devices.  Manipulating the topological properties of such systems has historically relied on controlling bulk properties of the system, which restricts the tunability of such systems.  For armchair graphene nano-ribbons, rather than manipulating the bulk, we show how modifying the dynamics at the ribbon edges alone can directly affect the bulk energy gap. This dependence is dominant in the presence of interactions typical for these system, and can induce topological phase transitions. Through the bulk/boundary correspondence we can localize states within a single width ribbon.  This opens the door to the development of quantum topological dots, quantum spin chains, spin qubits, and other fascinating systems relevant for quantum simulation.  Finally, we provide, for the first time, scaling relationships between the bulk gap and edge hopping parameters.
\end{abstract}

\keywords{Armchair graphene nanoribbons; Bulk energy gap; Full edge control; Topological phases}

\maketitle

The electronic energy gap is a fundamental energy scale of a solid, governing charge transport, thermal carrier activation and the stability of electronic phases, to name a few examples~\cite{AshcroftMermin1976,Kittel1996,chaves2020bandgap,HasanKane2010RMP,QiZhang2011RMP}. In a single-particle description, the intrinsic bulk energy gap of a solid-state system is defined as $E_{\mathrm{g}}= E_{\mathrm{CBM}}-E_{\mathrm{VBM}}$, where $E_{\mathrm{CBM}}$ and $E_{\mathrm{VBM}}$ denote the conduction-band minimum and valence-band maximum, respectively. In quasi-1D and quasi-2D materials with physical boundaries, a bulk gap, which separates extended valence- and conduction-band states, is governed predominantly by the interior bulk region. The contribution of the boundaries to the bulk gap is only through the small probability weight of those extended states within the boundary region.

%the energy gap can be generally classified into two groups, a boundary-involved gap and a purely bulk-defined gap. A boundary-involved gap separates two boundary states or one boundary state from an extended bulk state, which is highly sensitive to the variations at the boundaries such as boundary hopping, termination, hybridization and so on. Boundary engineering can consequently produce marked changes in a boundary-involved gap without comparably modifying the intrinsic bulk spectrum. By contrast, a purely bulk gap separates extended valence- and conduction-band states and is governed predominantly by the interior Hamiltonian. The contribution of the local boundaries to the bulk gap is only through the small probability weight of those extended states within the boundary region. Therefore, such boundary contribution compared with the bulk gap can be roughly estimated as ratio of boundary measure $S$ to system volume $V$, i.e., $S/V\sim 1/L$ with $L$ the characteristic confinement length. This indicates that the boundary contribution becomes asymptotically negligible as the characteristic confinement length of the system increases. 

\begin{figure}[htbp]
\includegraphics[width=\textwidth,height=0.65\textheight,keepaspectratio]{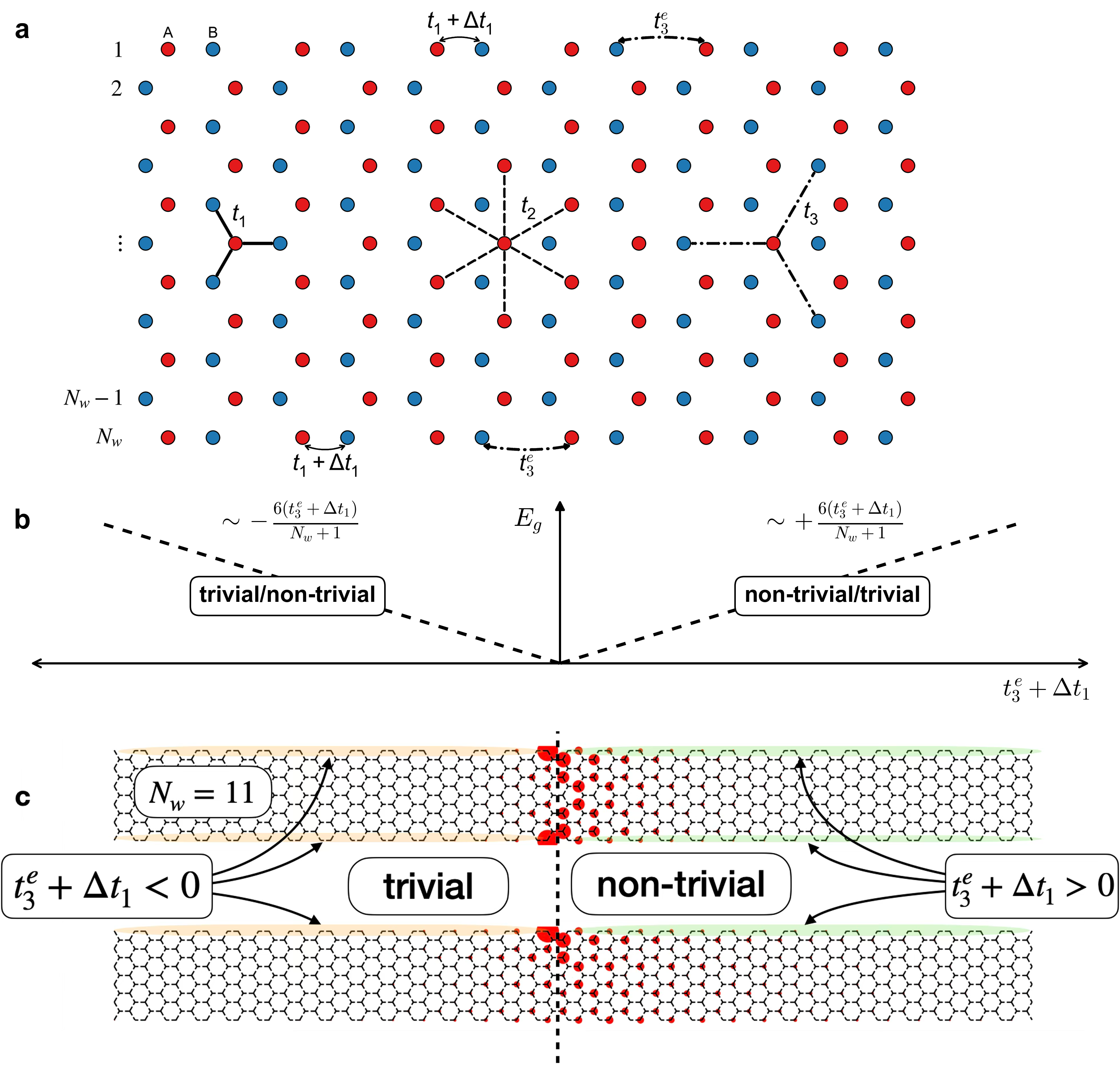}
\caption{(a) Lattice structure of AGNR with width $N_w$. Red (blue) circles denote A (B) sublattice sites. Representative first-, second- and third-NN hoppings, $t_1$, $t_2$ and $t_3$, are shown by solid, dashed and dash-dotted lines, respectively. The curved double-headed arrows denote the hoppings at the edges $t_1+\Delta t_1$ and $t^e_3$. (b) Dependence of gap $E_g$ on the parameter combination $t^e_3+\Delta t_1$ for width $N_w=3p+2$. A sign change of $t^e_3+\Delta t_1$ drives a topological phase transition between two distinct phases. (c) Localization of states, shown as red density circles, due to the junction of two topologically distinct equal-width ribbons.  The black arrows show where and how the edge dynamics are modified for the two example ribbons, and the vertical dashed line demarcates the topologically distinct sections. Here, we choose $N_w=11$ with the left region trivial while the right region non-trivial.\label{fig:lattice gap topology}}
%The presence of $U$ and $t_2$ will change slightly the location where the gap closes on the $x$-axis, but the qualitative features remain the same.
\end{figure}

Contrary to the prevailing view that in the bulk gap scenario the dynamics of the bulk dictates gap formation and dominates over the boundary contribution, we find that for armchair graphene nanoribbons (AGNRs) with width $N_w=3p+2$ for $p\in\mathbb{Z}\ge 0$, the bulk gap can be fully controlled by the hoppings \emph{solely at the boundaries} defining the width of the ribbon.  This full control of the bulk gap via boundary dynamics holds within the presence of interactions expected for these systems~\cite{magda2014room}, and for any width $N_w$ in this family.  This is the main finding of our work, and its significance can not be understated since the properties of the ribbons' bulk, e.g. thermal and electrical conductivity, can now be manipulated by simply modifying the boundary dynamics of the ribbon.  %Such modifications have recently been experimentally demonstrated~\cite{}.  
Such modifications could be made manifest during the synthetic manufacturing of ribbons~\cite{Li:2018synthetic,Zhang:2026bottom}, via careful hydrogenation of edge bonds~\cite{Lawrence:2022hydrogenation}, through directly tunable systems constructed with scanning tunneling microscopes~\cite{Drost:2017stm,Huda:2020stm}, or modeled from ion-trap systems~\cite{Meier:2016ions} and photonic lattices~\cite{Mukherjee:2017photonic,On:2024photonic}.
Furthermore, by careful engineering of the edge dynamics, one now has the ability to tune the ribbon with the same width into topologically distinct phases, and induce low-energy localized states between topologically distinct regions.  Figure~\ref{fig:lattice gap topology} demonstrates this scenario. This opens routes to engineering 1D band structures~\cite{rizzo18} and realizing quantum-device architectures, including qubits, topological quantum dots and artificial quantum chains—such as Su–Schrieffer–Heeger lattices~\cite{Groning:2018xdi,Wang:2026ahy}, as well as versatile platforms for quantum simulation.

As shown in Fig.~\ref{fig:lattice gap topology}(a), AGNRs are quasi-1D strips of graphene with armchair-edge terminations.  Depending on their width, they can have sizeable energy gaps and distinct topological states, making them already promising building blocks for nanoscale electronic and optoelectronic devices. AGNRs naturally fall into three families: $N_w=3p$, $3p+1$ and $3p+2$ where only the $N_w=3p+2$ family is predicted to be gapless in the simplest nearest-neighbor (NN) tight-binding model~\cite{Wakabayashi_2010}. % but opens a small finite gap when structural distortions at the edges or long-range hoppings are included. 
Bottom-up synthesis has enabled the experimental realization of AGNRs with atomic-scale control over their width and edge structure~\cite{Zhang:2026bottom}. In this work we will only consider the $N_w=3p+2$ family, though we note that our formalism described below can be applied to the other families as well.

\subsection*{Hamiltonian of AGNRs}
To demonstrate full edge control of the bulk gap, we consider the following Hamiltonian applied to the AGNRs,
\begin{equation}\label{eqn:hubbard H}
    H = \left( \sum_{\langle{i,j}\rangle,s}t_1 + \sum_{{\langle\langle{i,j}\rangle\rangle},s}t_2 + \sum_{{{{\langle\langle\langle{i,j}\rangle\rangle\rangle}}},s}t_3 \right) (c^\dagger_{i,s} c_{j,s}^{} + \mathrm{H.c.})-\frac{U}{2}\sum_i(n_{i,\uparrow}-n_{i,\downarrow})^2\ ,
\end{equation}
where $c^\dagger_{i,s}$ $(c_{i,s})$ creates (annihilates) an electron in the $p_z$ orbital at site $i$, $s=\uparrow,\downarrow$, and the number densities are $n_{i,s}=c^\dag_{i,s}c^{}_{i,s}$. Here $\langle\rangle$, $\langle\langle\rangle\rangle$, and $\langle\langle\langle\rangle\rangle\rangle$ denote first-, second-, and third-NN hopping, respectively. The corresponding hopping amplitudes are $t_1$, $t_2$ and $t_3$. The form of the Hubbard interaction $U$ in Eq.~\eqref{eqn:hubbard H} ensures that the half-filled state is the global ground state. In this work we consider the parameters $t_2$, $t_3$, and $U$ \emph{relative} to $t_1$, thereby making them dimensionless. For a sense of scale, typical \emph{dimensionful} values for these parameters are $t_1=-2.88$~eV, $t_2=0.22$~eV, and $t_3=-0.25$~eV~\cite{PhysRevLett.130.026401} and it is expected that $U/|t_1|<3.8$~\cite{magda2014room,Ostmeyer:2020uov,Ostmeyer:2021efs}.  
\begin{figure}[htbp]
\includegraphics[width=\textwidth,height=0.65\textheight,keepaspectratio]{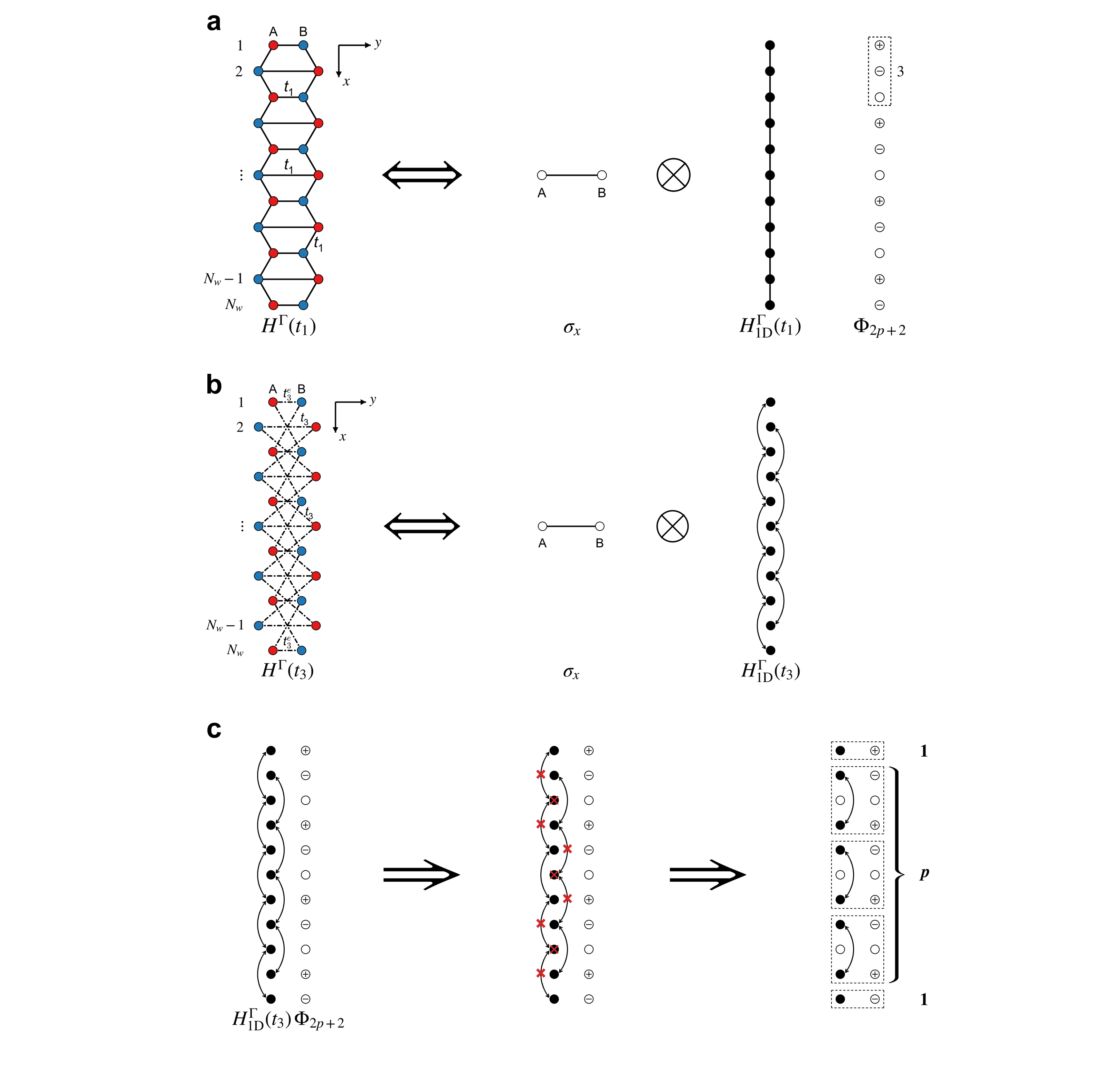}
\caption{One-dimensional mapping of Hamiltonian of AGNR unit cell at $\Gamma$ point. $H^\Gamma(t_i)=\sigma_x\otimes H^\Gamma_{\mathrm{1D}}(t_i)$
for $i=1,3$, where $\sigma_x$ acts in sublattice space and $H^\Gamma_{\mathrm{1D}}(t_i)$ describes an $N_w$-site 1D chain with on-site energy $t_i$ (black circles). (a) The solid lines denote the first NN hopping $t_1$. The zero-energy state $\Phi_{2p+2}$ exhibits the period-three amplitude pattern $(+,-,0)$, where $\pm$ denote amplitudes of equal magnitude and opposite sign, unlabelled open circles denote zero amplitude and the dashed box highlights one period. (b) Third NN hoppings in the unit cell are shown as dash-dotted lines and second NN hoppings $t_3$ in the chain are labelled by curved double-headed arrows. Top and bottom black circles in the chain represent the on-site energy $t^e_3$. (c) Decomposition of $H^{\Gamma}_{\rm 1D}(t_3)\Phi_{2p+2}$. Red crosses mark removed on-site and hopping terms associated with zero-amplitude sites. Dashed boxes identify two edge contributions and $p$ equivalent bulk contributions.}
\label{fig: t1 t3 gamma point}
\end{figure}

\subsection*{Energy gap opening at the \texorpdfstring{$\Gamma$}{Gamma} point}
We now systematically show the dependence of the gap on these different parameters in Eq.~\eqref{eqn:hubbard H}. When $U=0$, instead of working directly with Eq.~\eqref{eqn:hubbard H} we develop a new computational framework for calculating the energy gap of AGNRs by mapping the tight-binding Hamiltonian of a single unit cell at the symmetric $\Gamma$ point to the direct product of a Pauli matrix acting on the sublattice space and a tight-binding Hamiltonian of a finite 1D chain. This framework is completely analogous to solving the original tight-binding Hamiltonian at this specific $\Gamma$ point, and has the added benefit that we can directly calculate the topological invariant of the system due to mirror and/or inversion symmetry.  We can also apply our formalism at the symmetric $X$ point, as shown in the Methods section.  As an example, we first consider the dominant term in the tight-binding Hamiltonian with the first NN hopping parameters $t_1$. As shown in Fig.~\ref{fig: t1 t3 gamma point}(a), the Hamiltonian of the unit cell of the $N_w$-atom wide armchair ribbon with hoppings $t_1$ at the $\Gamma$ point, $H^{\Gamma}(t_1)$, is equivalent to the direct product of a Hamiltonian of an $N_w$-site 1D chain $H^{\Gamma}_{\rm 1D}(t_1)$ and a Pauli matrix $\sigma_x$, i.e., $H^{\Gamma}(t_1)=H^{\Gamma}_{\rm 1D}(t_1)\bigotimes\sigma_x$ with $\sigma$ the Pauli matrix of sublattice degree of freedom. The $N_w$-site 1D chain $H^{\Gamma}_{\rm 1D}(t_1)$ depends of the on-site energy $t_1$ labelled by black circles and the first NN hopping $t_1$ labelled by the solid lines.  A thorough description of $H^{\Gamma}_{\rm 1D}(t_1)$ is provided in Sec.~\ref{sect:A1} of the Supplementary Information. The on-site energy $t_1$ and the first NN hopping $t_1$ in the 1D chain originate from the first NN hopping along and across the armchair ribbon, respectively. Our formalism correctly predicts that the AGNR with $N_w=3p+2$ is gapless due to the presence of two zero-energy degenerate eigenstates, as shown in the Methods section. We concentrate on these two degenerate eigenstates, which we label
$\Phi_{\pm,2p+2}=\Phi_{2p+2}\bigotimes \ket{\pm}$ with $\ket{\pm}=\frac{1}{\sqrt{2}}(1,\pm 1)^T$, since they play the dominant role in determining the energy gap. Note that $\Phi_{2p+2}$ and $\ket{\pm}$ are the eigenstate of $H^{\Gamma}_{\rm 1D}(t_1)$ and $\sigma_x$, respectively. These two zero-energy wavefunctions exhibit the period-three pattern $(+,-,0)$ along the 1D chain with $\pm$ amplitudes of equal magnitude and opposite sign as shown in Fig.~\ref{fig: t1 t3 gamma point}(a). This unique property of the wavefunction will greatly simplify the calculations on the energy gap opening shown below, and is ultimately the reason that allows for control of the bulk energy gap via the edge.

We now consider the long-range $t_3$ hoppings. As shown in Fig.~\ref{fig: t1 t3 gamma point}(b), the Hamiltonian of the unit cell of the $N_w$-atom wide armchair nanoribbon with third NN hoppings $t_3$ at the $\Gamma$ point, $H^{\Gamma}(t_3)$, is equivalent to the direct product of a Hamiltonian of an $N_w$-site 1D chain $H^{\Gamma}_{\rm 1D}(t_3)$ and a Pauli matrix $\sigma_x$, i.e., $H^{\Gamma}(t_3)=H^{\Gamma}_{\rm 1D}(t_3)\bigotimes\sigma_x$. The $N_w$-site 1D chain $H^{\Gamma}_{\rm 1D}(t_3)$ includes on-site energy $t_3$ in the bulk region and $t^e_3$ at the edges labelled by black circles and the second NN hopping $t_3$ labelled by the curved double-headed arrows, with more details in Sec.~\ref{sect:A1} of the Supplementary Information. Here, the on-site energy and the second NN hopping in the 1D chain originate from the third NN hoppings along and across the ribbon, respectively. For the two zero-energy states, we have $H^{\Gamma}(t_3)\Phi_{\pm,2p+2}=H^{\Gamma}_{\rm 1D}(t_3)\Phi_{2p+2}\bigotimes \sigma_x\ket{\pm}$. In Fig.~\ref{fig: t1 t3 gamma point}(c), we show the decomposition of $H^{\Gamma}_{\rm 1D}(t_3)\Phi_{2p+2}$ based on the period-three pattern $(+,-,0)$ of $\Phi_{2p+2}$. After removing all the on-site and hopping terms associated with zero-amplitude sites, $H^{\Gamma}_{\rm 1D}(t_3)\Phi_{2p+2}$ can be decoupled into $p+2$ parts indicated by the dashed boxes, i.e., one at the top edge with $t^e_3\phi_{2p+2}(x_1)$, one at the bottom edge with $t^e_3\phi_{2p+2}(x_{N_w})$, and $p$ equivalent parts in the bulk region with $t_3(\eta_0+\eta_x)\Big(\phi_{2p+2}(x_{3j-1}),\phi_{2p+2}(x_{3j+1})\Big)^T=t_3(\eta_0+\eta_x)C_{N_w}(-1,1)^T=0$. Here, $\phi_{2p+2}(x_i)(1\le i\le N_w)$ represents the amplitude of $\Phi_{2p+2}$ at site $x_i$ with more details in the Methods section, $\eta_0$ and $\eta_x$ denote the identity and $x$-Pauli matrices in the basis of two sites $\{x_{3j-1},x_{3j+1}\}(1\le j\le p)$, respectively. Interestingly, for all $p$ parts in the bulk region, the contribution due to the on-site energy $t_3\eta_0$ and the hopping $t_3\eta_x$ exactly cancel each other. This indicates that all $t_3$ terms in the bulk region has no effect on the two zero-energy states and therefore have no contribution to energy gap opening. In contrast, for the parts at both the top and bottom edges, the on-site energy $t^e_3$ alone gives rise to the energy gap 
\begin{displaymath}
E_g(t^e_3)=\frac{6|t^e_3|}{N_w+1}\ ,
\end{displaymath}
showing that the gap opens at $\mathcal{O}(t^e_3)$, with more details in the Methods section. We emphasize that this energy gap is entirely due to the third NN hopping $t_3$ \emph{at the edges} of armchair ribbons. We also note that this is the first prediction of such a scaling law of the energy gap as a function of ribbon width in AGNRs with $N_w=3p+2$. A similar method can be applied to calculate the effect of variation of the first NN hopping parameters at the edges $\Delta t_1$ due to the change of the inter-atomic distance between carbon atoms. One obtains the induced energy gap $E_g(\Delta t_1)=6|\Delta t_1|/(N_w+1)$, in agreement with previous studies~\cite{PhysRevLett.97.216803}.

The dependence of the energy gap on the long-range $t_2$ hoppings can be deduced from a similar mapping to a 1D chain, and we leave the details in this case for the Methods section.  We find that the influence of $t_2$ on the gap first occurs at order $\mathcal{O}(t_2^2)$, and there are two contributions.  The first is an overall irrelevant energy shift, and the second is a gap opening. Ignoring the constant energy shift, the gap dependence on $N_w>2$ and $t_2$ is 
\begin{displaymath}
E_g(t_2)=\frac{9(1+2p)t^2_2}{2|t_1|(N_w+1)^2}\ .
\end{displaymath}
For $N_w=2$ there is no such gap opening due to $t_2$. We note that our finding of a gap opening due to $t_2$ is new and contrary to previous statements found in the literature~\cite{PhysRevLett.130.026401}.  We have verified our findings with direct numerical calculations as shown in the Methods section.

We now consider the effect of the Hubbard interaction on the gap opening. Again, the details of our derivations are given in Sec.~\ref{sect:A3} of the Supplementary Information and we only present the salient results here. We employ degenerate perturbation theory in $U$\footnote{The actual expansion parameter is $U/2$.}.  
To calculate the gap, we calculate the shift in energy $E_{Q=0}$ of the half-filled neutral $Q=0$ ground state as well as the corresponding shift of $E_{Q=1}$ due to the single-particle $Q=1$ state above half-filling.  The gap is then given by $E_g(U)=2(E_{1}-E_{0})$.  As is well known~\cite{PhysRevB.50.6939,Gebhard:2003} there is no gap induced at leading order in $U$ but rather an irrelevant constant energy shift. The gap first opens at order $\mathcal{O}(U^2)$.  At this order we find that the ground state is a singlet in spin, in agreement with Lieb's condition~\cite{PhysRevLett.62.1201,PhysRevLett.62.1927.5,Yang1989EtaPairing,YangZhang1990SO4}.  In this case, we are not able to obtain an analytic scaling law for the gap due to interactions $U$, but we empirically observe that 
\begin{displaymath}
    E_g(U)\propto \frac{U^2}{4N_w}\ .
\end{displaymath}

\begin{figure}[tbp]
\includegraphics[width=1.0\columnwidth]{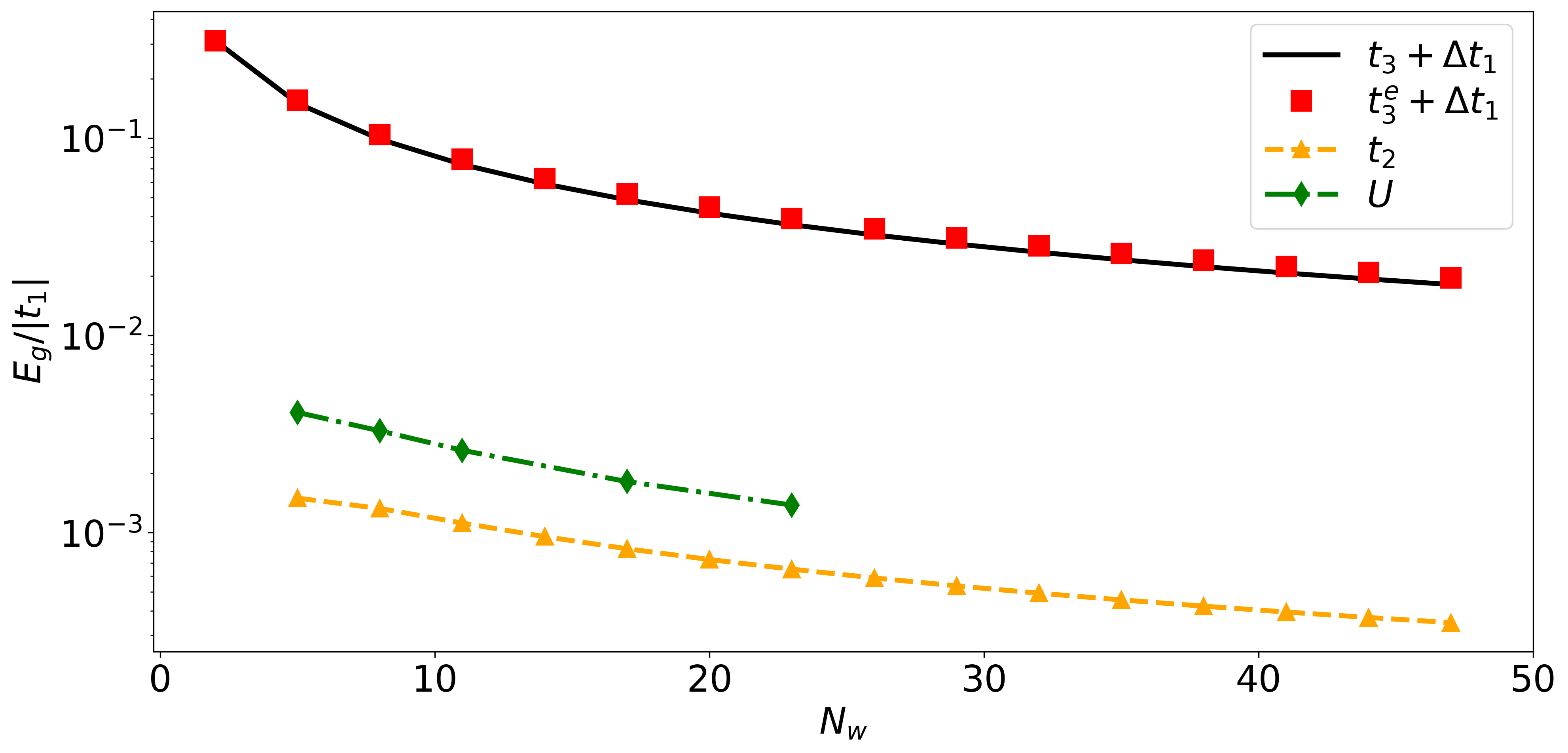}
\caption{Energy gap $E_g$ as a function of armchair nanoribbon width $N_w=3p+2$. The solid line, dashed line with triangles and dash-dotted line with diamonds denote the energy gap due to $t_3+\Delta t_1$, $t_2$ and $U$, respectively. The red squares stand for the analytical result from $E_g=(6|t^e_3+\Delta t_1|)/(N_w+1)$ due to the hopping $t^e_3$ and $\Delta t_1$ at the edges. Here, $t_1=-2.88\ $eV, $t_3=-0.25\ $eV, $t_2=0.22\ $eV, $\Delta t_1=-0.2\ $eV and $U=|t_1|$.}
\label{fig: energy gap}
\end{figure}
Based on the above analytical derivations, the resulting total energy gap due to hoppings is $E_g=|(6(t^e_3+\Delta t_1))/(N_w+1)-(9(1+2p)t^2_2)/(2t_1(N_w+1)^2)|$ for $N_w>2$ and $E_g=|(6(t^e_3+\Delta t_1))/(N_w+1)|$ for $N_w=2$.  In Fig.~\ref{fig: energy gap}, we show a breakdown of the contribution of the gap due to edge hoppings $t^e_3+\Delta t_1$ and $t_2$ and the Hubbard interaction $U$, as a function of ribbon width $N_w$.  The specific values of these parameters are found in the accompanying caption.  As can be seen from the figure, the induced gap due to both $t_2$ and $U$ are at least an order of magnitude smaller than that induced by the combination $t^e_3+\Delta t_1$ at the edges.  As a result, the total energy gap is effectively determined by $E_g=|(6(t^e_3+\Delta t_1))/(N_w+1)|$, which is the basis for our statement of having full edge control of the bulk gap. Figure~\ref{fig:lattice gap topology}(b) shows this exact energy gap dependence.

\subsection*{\texorpdfstring{$Z_2$}{Z2} topological invariant}
In gapped AGNRs with mirror and/or inversion symmetries, the topological invariant $Z_2$ is calculated by $(-1)^{Z_2}=\prod_n[P_n(\Gamma)P_n(X)]$~\cite{PhysRevLett.119.076401,PhysRevLett.130.026401}, where $P_n(\Gamma)$ and $P_n(X)$ are the parity of the eigenstates in the $n$-th band at the $\Gamma$ and $X$ points of the 1D Brillouin zone, respectively. This product is taken over all occupied bands. The invariant $Z_2=1$ $(Z_2=0)$ denotes a nontrivial (trivial) band topology. As stated earlier, our formalism allows us to determine the parities at both $\Gamma$ and $X$ points.  Since the contribution due to $t_2$ and $U$ are negligible, we completely ignore them in our analysis.  We find that at the $X$ point there can never be a band inversion due to changes in $t^e_3+\Delta t_1$, while at the $\Gamma$ point there will always occur a band inversion between the two near-zero-energy states that determine the band gap when the sign of $t^e_3+\Delta t_1$ changes.  This means that the topological phases are always distinct when the sign of $t^e_3+\Delta t_1$ changes, thus giving different $Z_2$ invariant.  In the Methods section we enumerate the exact invariants and their dependence on $N_w$ and symmetry and different unit cells.

% At $\Gamma$ point, under mirror symmetry $\mathcal{M}_y:\ (x,y)\mapsto(x,-y)$ with $y$ the periodic direction, the states $\Phi_{+,m}$ $(\Phi_{-,m})$ have even (odd) parity since $\Phi_m$ remain invariant and $\ket{\pm}$ becomes $\pm\ket{\pm}$. Then, the number of the occupied states with odd parity is determined by $E_{-,m}<0$. Note that the energy shift and negligible energy gap due to the hopping $t_2$ are not taken into account here. One obtains $\prod_nP_n(\Gamma)=(-1)^{\lfloor \frac{N_w}{3} \rfloor}$ for $t_3+\Delta t_1<0$. Under inversion symmetry $\mathcal{I}:\ (x,y)\mapsto(-x,-y)$, $\Phi_m$ becomes $(-1)^{m+1}\Phi_m$ and $\ket{\pm}$ becomes $\pm\ket{\pm}$, which indicates that the states $\Phi_{+,m}$ with even $m$ and $\Phi_{-,m}$ with odd $m$ have odd parity. As a result, the number of the occupied states with odd parity is determined by $E_{+,m}<0$ with even $m$ and $E_{-,m}<0$ with odd $m$, leading to $\prod_nP_n(\Gamma)=(-1)^{\lfloor \frac{N_w-1}{2} \rfloor+1}$ for $t_3+\Delta t_1<0$. Together with the parity of the wavefunctions at $X$ point obtained in App.~\ref{sect:A3}, we list in Table~\ref{tab:topological_invariant} the topological invariant for different unit-cell types associated with mirror symmetry $\mathcal{M}_y$ and inversion symmetry $\mathcal{I}$, for AGNRs belonging to the $N_w=3p+2$ family, in the case of $t_3+\Delta t_1<0$.

\subsection*{Discussion and outlook}
Taken as a whole, our findings presented above provide the ability to modify the bulk gap by manipulating edge dynamics, and this in turn provides a handle on controlling bulk properties of the armchair ribbon.  Since the topological phases are distinct and depend on the sign of $t^e_3+\Delta t_1$, one could, for example, drive a system through a topological phase transition (i.e. gap closing and reopening) by purely changing the sign of $t^e_3+\Delta t_1$ since these parameters are dominant compared to $t_2$ and $U$.  Depending on the width $N_w$, one could tune the system from a trivial insulator to a non-trivial insulator, or vice versa.  \autoref{fig:lattice gap topology}(c) shows this scenario for the specific case $N_w=11$. 

Via the bulk/edge correspondence~\cite{HasanKane2010RMP,QiZhang2011RMP}, one can induce electronic localized states by joining ribbons with different topological invariants.  This has been done with hybrid ribbons of different widths and invariants~\cite{Groning:2018xdi,rizzo18,Luu:2022efw,Ostmeyer:2024dhw}.  Our work here shows that we can achieve similar localizations with a ribbon of a single width, but where sections of the ribbon have altered edge dynamics and thus differing invariants.  In Fig.~\ref{fig:lattice gap topology}(c) we show two cases where the low-energy electronic state is localized at the intersection of two topologically distinct equal-width armchair ribbon with $N_w=11$.  What is shown is the wavefunction densities, whereby the radius of the circle is proportional to the density.  In one case both edges of the ribbon are modified to incur different invariants, while in the other only one edge of the ribbon is modified. The edge modifications shown in Fig.~\ref{fig:lattice gap topology}(c) can be used as building blocks to construct arrays of localized states.  By alternating the edge dynamics, one can engineer localizations at specific points along the length of the ribbon, as well as at specific sides of the ribbon.  Artificial quantum chains can be atomically manufactured to simulate, for example, the SSH model~\cite{Wang:2026ahy,Groning:2018xdi}.  Qubits and 1D band structures are also conceivable in this manner~\cite{rizzo18,wang2021graphene}.

\subsection*{Conclusions}
In summary we have demonstrated how the bulk energy gap in armchair nanoribbons with width $N_w=3p+2$ with $p\in \mathbb{Z}\ge 0$ depends on the hopping parameters $t_3$, $\Delta t_1$ and $t_2$, and on the Hubbard interaction $U$ to second order in perturbation theory. We find that the dominant contribution to the bulk energy gap comes from the combination $t^e_3+\Delta t_1$ at the edges, with $t_2$ and $U$ contribution at least an order of magnitude smaller. We provide, for the first time, scaling laws for the bulk gap due to $t_3$, $\Delta t_1$, and $t_2$. Strikingly, tuning $t^e_3$ and $\Delta t_1$ \emph{exclusively at the ribbon edges} provides full control over the bulk energy gap, which is contrary to the prevailing view that the bulk energy gap defined by extended bulk states is dominated by the interior bulk region. Such full edge control of energy gap potentially provides a simple manner to manipulating the bulk properties of the material since only edge dynamics must be modified.
Finally, we have determined the topological phases of the ribbon to be distinct and dependent on the sign of $t^e_3+\Delta t_1$. This enables the design of equal-width ribbons with distinct topological invariants, offering a route to engineering 1D band structures and potential platforms for qubits, artificial quantum chains and other quantum devices. Future work will explore whether complete boundary control of bulk energy gap extends to other solid-state systems with physical boundaries.

\section*{Data availability}
The data pertaining to this study are available from the corresponding author upon reasonable request.

\begin{acknowledgments}
	This work was funded in part by the Deutsche
	Forschungsgemeinschaft (DFG, German Research Foundation) as part of the CRC 1639 NuMeriQS–project no.~511713970.
	We gratefully acknowledge the computing time granted by the JARA Vergabegremium and provided on the JARA Partition part of the supercomputer JURECA at Forschungszentrum Jülich~\cite{jureca-2021}.  The work of UGM was supported in part by the CAS President's International Fellowship Initiative (PIFI) under Grant No. 2025PD0022. 
\end{acknowledgments}

\section*{Author contributions}
L.W. conceived and initiated the project. L.W. performed the calculations on the energy gap opening and topological invariant using the tight-binding Hamiltonian. T.L. performed second-order degenerate perturbation theory calculations on the Hubbard interaction, contributed to localization of states at the junction of two topologically distinct equal-width ribbons and generated various figures for the manuscript. U.-G.~M. performed consistency checks of calculations, provided guidance throughout the research project and aided in construction and layout of manuscript, as well as supervising the layout of figures. All authors contributed to the writing, editing and review of this work. 

\section*{Competing interests}
The authors declare no competing interests.

\bibliography{references}

\clearpage
\section*{Methods}
% \subsection*{Model}
% As shown in Fig.~\ref{fig: unit cell t1 gamma point}(a), armchair graphene nanoribbon (AGNR) is a quasi-1D strip of graphene with armchair-edge termination. $N_w$ denotes the ribbon width, A and B represent the sublattice sites and $t_1$, $t_2$ and $t_3$ are the first, second, and third nearest-neighbor hopping parameters, respectively. The tight-binding Hamiltonian with all these hopping parameters as well as the Hubbard interaction for AGNRs is given by
% \begin{equation}\label{eqn:hubbard H}
%     H = \left( \sum_{\langle{i,j}\rangle,s}t_1 + \sum_{{\langle\langle{i,j}\rangle\rangle},s}t_2 + \sum_{{{{\langle\langle\langle{i,j}\rangle\rangle\rangle}}},s}t_3 \right) (c^\dagger_{i,s} c_{j,s}^{} + \mathrm{H.c.})-\frac{U}{2}\sum_i(n_{i,\uparrow}-n_{i,\downarrow})^2\ ,
% \end{equation}
% where $c^\dagger_{i,s}$ $(c_{i,s})$ creates (annihilates) an electron in the $p_z$ orbital at site $i$, $s=\uparrow,\downarrow$, and the number densities are $n_{i,s}=c^\dag_{i,s}c^{}_{i,s}$. The form of the Hubbard interaction $U$ in Eq.~\eqref{eqn:hubbard H} ensures that the half-filled state is the global ground state. In this work we consider the parameters $t_2$, $t_3$, and $U$ \emph{relative} to $t_1$, thereby making them dimensionless. For a sense of scale, typical \emph{dimensionful} values for these parameters are $t_1=-2.88$~eV, $t_2=0.22$~eV, and $t_3=-0.25$~eV~\cite{}. %In this work, we start with a spinless system in the non-interacting limit by setting $U=0$ in Eq.~\eqref{eqn:hubbard H} and then take into account the effect of Hubbard interaction.

\subsection*{Energy gap opening at the \texorpdfstring{$\Gamma$}{Gamma} point}
For the first NN $t_1$ hopping, the energy levels and eigenstates of $H^{\Gamma}_{\rm 1D}(t_1)$ are $E_m=t_1+2t_1\cos(\frac{\pi m}{N_w+1})$ and $\Phi_m=\Big(\phi_m(x_1),\phi_m(x_2),\ldots,\phi_m(x_{N_w})\Big)^T$ with $\phi_m(x_i)=\sqrt{\frac{2}{N_w+1}}\sin(\frac{\pi m x_i}{N_w+1})$, where $m=1,2,...,N_w$ is the quantum number that labels the eigenstates, $x_i=1,2,...,N_w$ is the coordinate of the $i$-th site along the chain. As a result, the eigenvalues and eigenstates of $H^{\Gamma}(t_1)$ are $E_{\pm,m}=\pm E_m$ and $\Phi_{\pm,m}=\Phi_m\bigotimes \ket{\pm}$ with $\ket{\pm}=\frac{1}{\sqrt{2}}(1,\pm 1)^T$ correspondingly. This gives rise to the gap closing condition $E_m=0$ with $m=2(N_w+1)/3$, which requires $N_w=3p+2$ with $p$ a non-negative integer, in agreement with previous studies~\cite{PhysRevLett.97.216803, PhysRevLett.130.026401}. We focus on the armchair nanoribbons with the width $N_w=3p+2$. Specially, for $m=2p+2$, there are two degenerate zero-energy eigenstates 
with $\Phi_{\pm,2p+2}=\Phi_{2p+2}\bigotimes \ket{\pm}$ where $\Phi_{2p+2}=C_{N_w}\Big(f(x_1),f(x_2),\ldots,f(x_{N_w})\Big)^T$ with 
$C_{N_w}=\sqrt{\frac{2}{N_w+1}}\sin(\frac{2\pi}{3})$ and 
\begin{equation}\label{eqn: phi}
f(x_i)=
\begin{cases}
1, & x_i \bmod 3 = 1,\\
-1, & x_i \bmod 3 = 2,\\
0, & x_i \bmod 3 = 0.
\end{cases}
\end{equation}
It is evident that these two zero-energy wavefunctions exhibit the period-three pattern $(+,-,0)$ along the 1D chain with $\pm$ amplitudes of equal magnitude and opposite sign as shown in Fig.~\ref{fig: t1 t3 gamma point}(a). This unique property of the wavefunction greatly simplifies the calculations of the energy gap opening. In the following, we focus on these two zero-energy states $\Phi_{\pm,2p+2}$ since they are the most relevant to determine the energy gap. 

\begin{figure}[htbp]
\includegraphics[width=1.0\columnwidth]{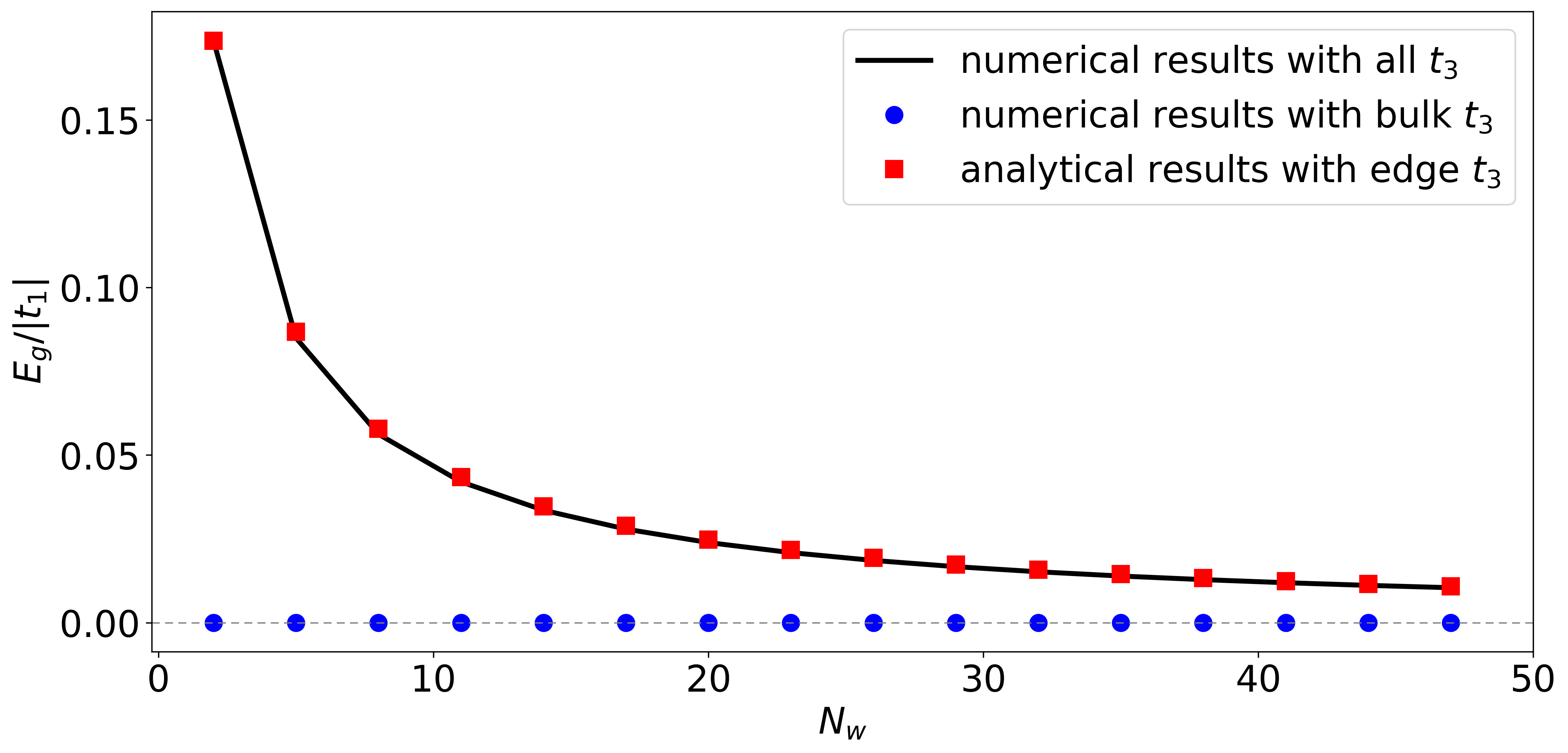}
\caption{Energy gap $E_g$ as a function of armchair nanoribbon width $N_w=3p+2$. The black solid line and the blue circles denote the numerical calculation from the tight-binding Hamiltonian with all the hoppings $t_3$ and $t_3$ in the bulk region, respectively. The red squares stand for the analytical result from $E_g=\frac{6|t^e_3|}{N_w+1}$ due to the hopping $t_3$ at the edges. Here, $t_1=-2.88\ $eV, $t_3=-0.25\ $eV, and $t_2=\Delta t_1=U=0$.}
\label{fig: energy gap t3}
\end{figure}

For long-range hoppings $t_3$, in the first order, the contribution to the two states $\Phi_{\pm,2p+2}$ is given by $E^{(1)}_{\pm,2p+2}(t_3)=\Phi^{\dagger}_{\pm,2p+2}H^{\Gamma}(t_3)\Phi_{\pm,2p+2}=\pm \Phi^{\dagger}_{2p+2}H^{\Gamma}_{\rm 1D}(t_3)\Phi_{2p+2}$ and therefore we have $E^{(1)}_{\pm,2p+2}(t_3)=\pm\left[\phi^{\dagger}_{2p+2}(x_1)t^e_3\phi_{2p+2}(x_1)+\phi^{\dagger}_{2p+2}(x_{N_w})t^e_3\phi_{2p+2}(x_{N_w})\right]$ that comes only from contributions at the edges. Since $\phi_{2p+2}(x_1)=-\phi_{2p+2}(x_{N_w})=\sqrt{\frac{2}{N_w+1}}\sin(\frac{2\pi}{3})$ by referring to Eq.~\eqref{eqn: phi}, we obtain $E^{(1)}_{\pm,2p+2}(t_3)=\pm \frac{3t^e_3}{N_w+1}$ and therefore the induced energy gap is given by $E_g(t^e_3)=|E^{(1)}_{+,2p+2}(t_3)-E^{(1)}_{-,2p+2}(t_3)|=\frac{6|t^e_3|}{N_w+1}$. We emphasize that such energy gap is entirely due to the third NN hopping $t_3$ \emph{at the edges} of armchair ribbons. %We also note that this is the first time to predict such scaling law of the energy gap as a function of width in armchair nanoribbons with the width $N_w=3p+2$. 
We have verified our analytical derivation with direct numerical calculations as shown in Fig.~\ref{fig: energy gap t3}. The numerically evaluated contribution of the bulk $t_3$ hopping to the energy gap vanishes identically, in exact agreement with our analytical derivation. Moreover, the gap due to the edge hopping $t_3^e$ reproduces the full numerical result including all $t_3$ hopping terms. A similar method can be applied to calculate the effect of variation of the first NN hopping parameters at the edges $\Delta t_1$ due to the change of the inter-atomic distance between carbon atoms. One obtains $E^{(1)}_{\pm,2p+2}(\Delta t_1)=\pm\frac{3\Delta t_1}{N_w+1}$ and the induced energy gap $E_g(\Delta t_1)=|E^{(1)}_{+,2p+2}(\Delta t_1)-E^{(1)}_{-,2p+2}(\Delta t_1)|=\frac{6|\Delta t_1|}{N_w+1}$, in agreement with~\cite{PhysRevLett.97.216803}.

\begin{figure}[htbp]
\includegraphics[width=1.0\columnwidth]{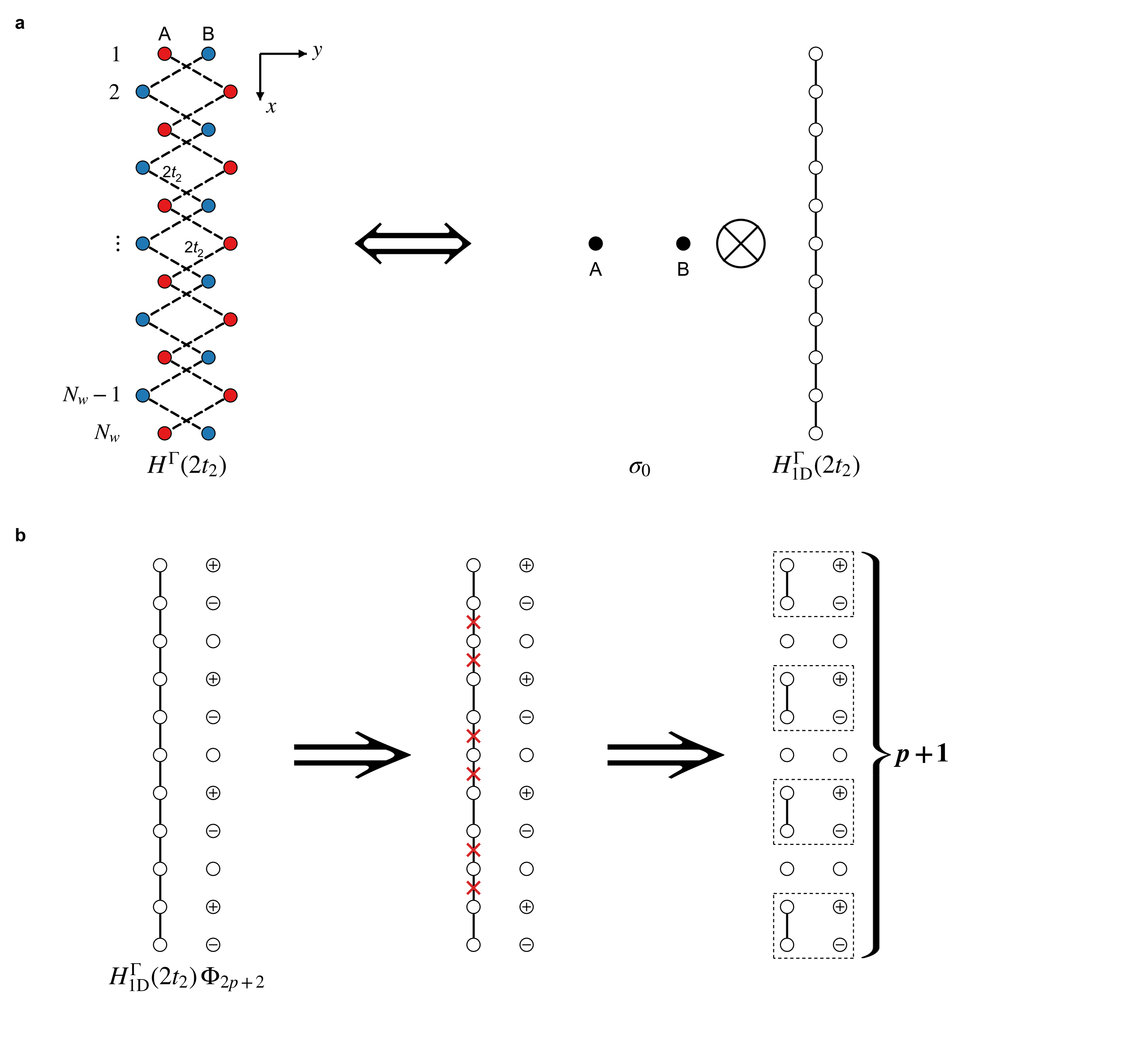}
\caption{(a) Mapping of Hamiltonian of AGNR unit cell with the second NN hopping $2t_2$ at $\Gamma$ point $H^{\Gamma}(2t_2)$ to direct product of $\sigma_0\bigotimes H^{\Gamma}_{\rm 1D}(2t_2)$, where $\sigma_0$ acts on sublattice space and $H^{\Gamma}_{\rm 1D}(2t_2)$ describes an $N_w$-site 1D chain. In the unit cell, the dashed lines denote the hopping $2t_2$. In 1D chain, the open circles and the solid lines represent zero on-site energy and the first NN hopping $2t_2$, respectively. (b) Decomposition of $H^{\Gamma}_{\rm 1D}(2t_2)\Phi_{2p+2}$ based on the period-three pattern $(+,-,0)$ of $\Phi_{2p+2}$. Red crosses indicate the hopping terms removed associated with zero-amplitude sites. This decouples $H^{\Gamma}_{\rm 1D}(2t_2)\Phi_{2p+2}$ into $p+1$ equivalent parts, as indicated by the dashed boxes.}
\label{fig: unit cell 2t2 gamma point}
\end{figure}

\begin{figure}[htbp]
\includegraphics[width=1.0\columnwidth]{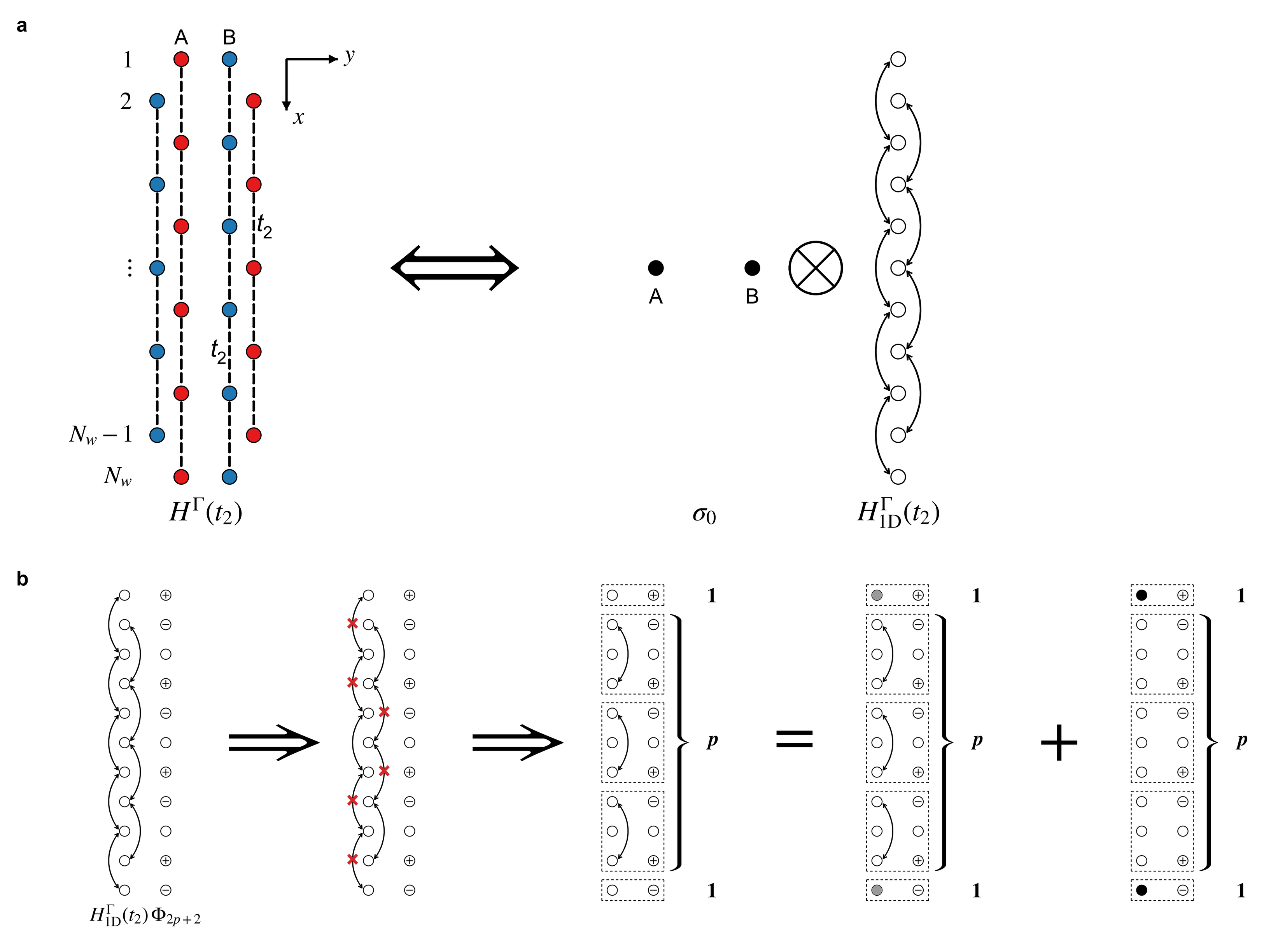}
\caption{(a) Mapping of Hamiltonian of AGNR unit cell with the second NN hopping $t_2$ at $\Gamma$ point $H^{\Gamma}(t_2)$ to direct product of $\sigma_0\bigotimes H^{\Gamma}_{\rm 1D}(t_2)$, where $\sigma_0$ acts on sublattice space and $H^{\Gamma}_{\rm 1D}(t_2)$ represents an $N_w$-site 1D chain. In the unit cell, the dashed lines denote the hopping $t_2$. In 1D chain, the open circles and the curved double-headed arrows represent zero on-site energy and the second NN hopping $t_2$, respectively. (b) Decomposition of $H^{\Gamma}_{\rm 1D}(t_2)\Phi_{2p+2}$ according to the period-three structure $(+,-,0)$ of $\Phi_{2p+2}$. Red crosses indicate the hopping terms removed associated with zero-amplitude sites. This decouples $H^{\Gamma}_{\rm 1D}(t_2)\Phi_{2p+2}$ into $p+2$ parts, i.e., one at the top (bottom) edge and $p$ equivalent bulk contributions, as indicated by the dashed boxes. The part at the top (bottom) edge further splits into two parts, i.e., one with on-site energy $-t_2$ labeled by the gray circles and another one with on-site energy $t_2$ indicated by the black circles.}
\label{fig: unit cell t2 gamma point}
\end{figure}

 For the second NN $t_2$ hopping, each site is connected to six neighbors in the lattice structure of AGNR [see Fig.~\ref{fig:lattice gap topology}(a)]. For the Hamiltonian of an AGNR unit cell at the $\Gamma$ point, these six hoppings are grouped as $2t_2+2t_2+t_2+t_2$ where the hopping $2t_2$ arises from two translationally equivalent hopping paths. The AGNR unit cell with the hoppings $2t_2$ and the hoppings $t_2$ are shown in Fig.~\ref{fig: unit cell 2t2 gamma point}(a) and Fig.~\ref{fig: unit cell t2 gamma point}(a), respectively. As presented in Fig.~\ref{fig: unit cell 2t2 gamma point}(a), the Hamiltonian of the unit cell of the $N_w$-atom wide armchair nanoribbon with the hoppings $2t_2$ $H^{\Gamma}(2t_2)$ is equivalent to direct product of a Hamiltonian of an $N_w$-site 1D chain $H^{\Gamma}_{\rm 1D}(2t_2)$ and an identity matrix $\sigma_0$, i.e., $H^{\Gamma}(2t_2)=H^{\Gamma}_{\rm 1D}(2t_2)\bigotimes\sigma_0$. The $N_w$-site 1D chain $H^{\Gamma}_{\rm 1D}(2t_2)$ only includes the first NN hopping $2t_2$ indicated by the solid lines, with more details in Sec.~\ref{sect:A1} of the Supplementary Information. For two zero-energy states, we have $H^{\Gamma}(2t_2)\Phi_{\pm,2p+2}=H^{\Gamma}_{\rm 1D}(2t_2)\Phi_{2p+2}\bigotimes \sigma_0\ket{\pm}$. In Fig.~\ref{fig: unit cell 2t2 gamma point}(b), we show the decomposition of $H^{\Gamma}_{\rm 1D}(2t_2)\Phi_{2p+2}$ based on the period-three pattern $(+,-,0)$ of $\Phi_{2p+2}$. After removing all the hopping terms associated with zero-amplitude sites, $H^{\Gamma}_{\rm 1D}(2t_2)\Phi_{2p+2}$ can be decoupled into $p+1$ equivalent parts indicated by the dashed boxes. For each part, $2t_2\eta_x\Big(\phi_{2p+2}(x_{3j+1}),\phi_{2p+2}(x_{3j+2})\Big)^T=2t_2\eta_x C_{N_w}(1,-1)^T=-2t_2\Big(\phi_{2p+2}(x_{3j+1}),\phi_{2p+2}(x_{3j+2})\Big)^T$ where $\eta_x$ represents the $x$-Pauli matrix in the basis of two sites $\{x_{3j+1},x_{3j+2}\}(0\le j\le p)$. This gives rise to $H^{\Gamma}_{\rm 1D}(2t_2)\Phi_{2p+2}=-2t_2\Phi_{2p+2}$, which indicates that $\Phi_{2p+2}$ is the eigenstate of $H^{\Gamma}_{\rm 1D}(2t_2)$ with the eigenvalue $-2t_2$. As a result, $H^{\Gamma}(2t_2)\Phi_{\pm,2p+2}=-2t_2\Phi_{\pm,2p+2}$ and therefore $E_{\pm,2p+2}(2t_2)=\Phi^{\dagger}_{\pm,2p+2}H^{\Gamma}(2t_2)\Phi_{\pm,2p+2}=-2t_2$. This shows that the hoppings $2t_2$ contribute to an overall energy shift. % of two zero-energy states. Such an energy shift can also be understood as follows. 
 Alternatively, it can be shown that $[H^{\Gamma}(t_1), H^{\Gamma}(2t_2)]=0$ and therefore these two Hamiltonians share the same eigenstates. Specifically, the two degenerate states $\Phi_{\pm,2p+2}$ are also eigenstates of $H^{\Gamma}(2t_2)$, and thus we have $E_{\pm,2p+2}(2t_2)=\Phi^{\dagger}_{\pm,2p+2}H^{\Gamma}(2t_2)\Phi_{\pm,2p+2}=\Phi^{\dagger}_{2p+2}H^{\Gamma}_{\rm 1D}(2t_2)\Phi_{2p+2}=-2t_2$. This exactly agrees with our derivation.

\begin{figure}[htbp]
\includegraphics[width=1.0\columnwidth]{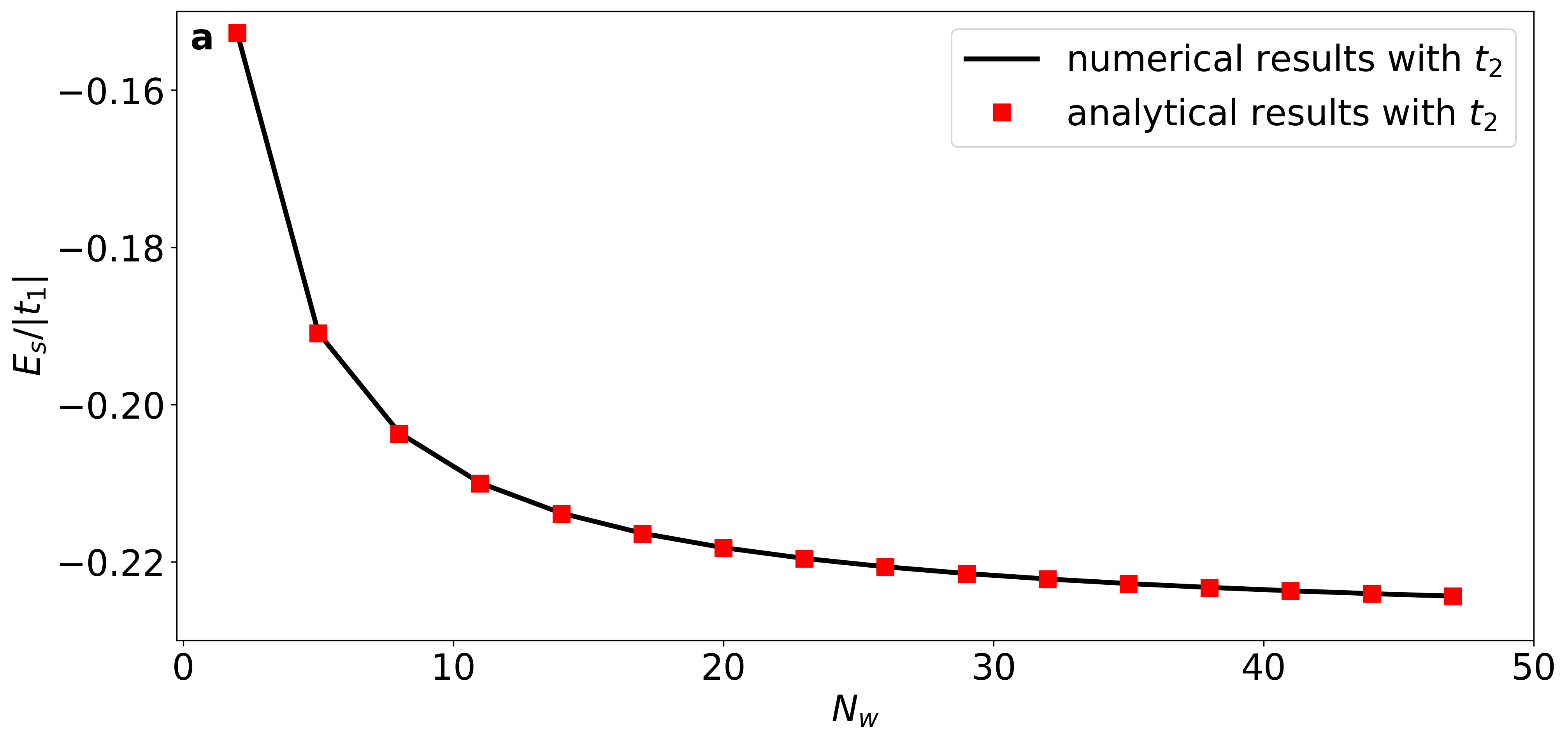}
\includegraphics[width=1.0\columnwidth]{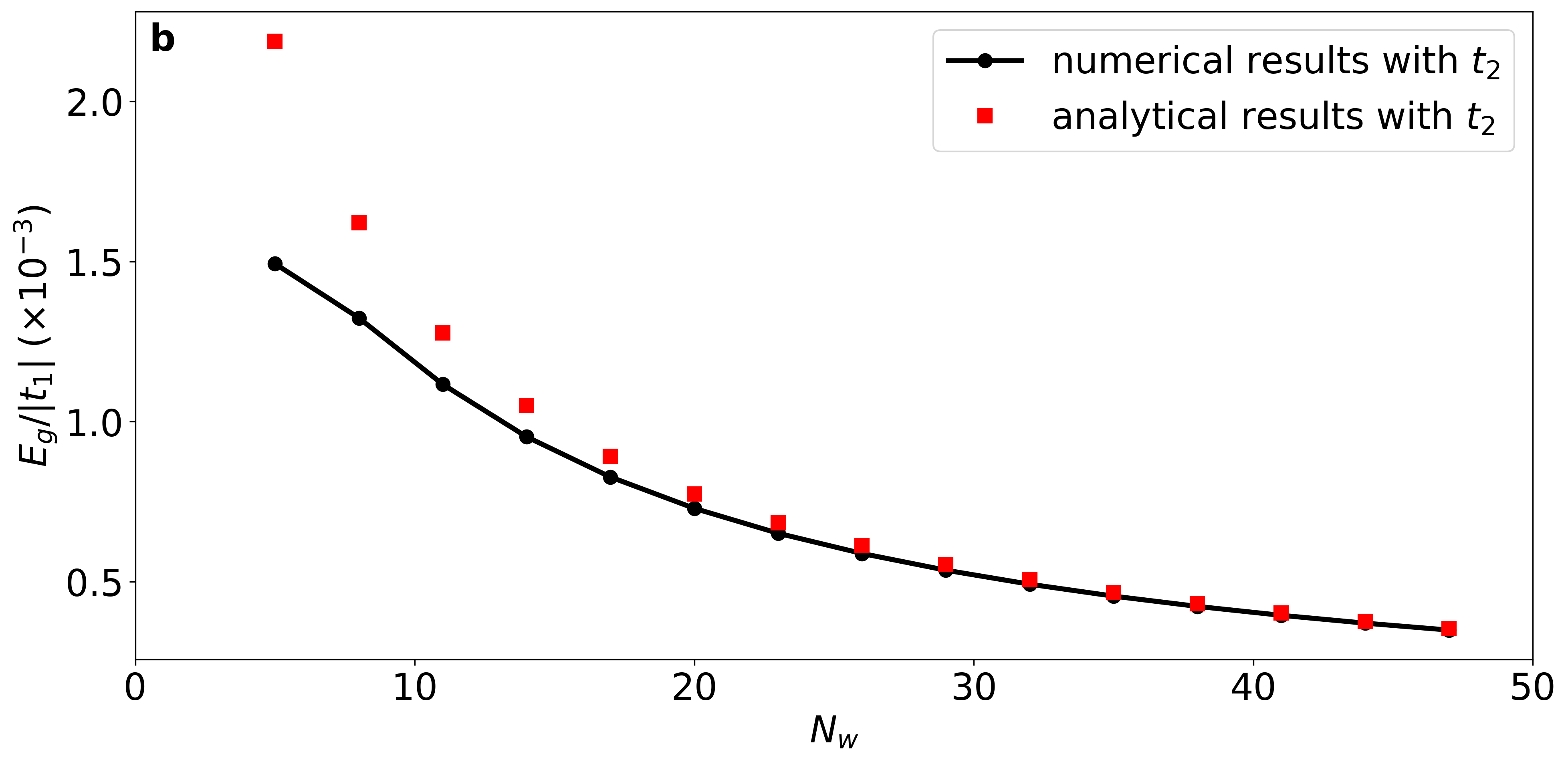}
\caption{(a) Energy shift $E_s$ as a function of armchair nanoribbon width $N_w$. The black solid line and the red squares represent the numerical calculation from the tight-binding Hamiltonian with all the hoppings $t_2$ and the analytical result from $E_s=-3t_2+\frac{3t_2}{N_w+1}$, respectively. (b) Energy gap $E_g$ as a function of armchair nanoribbon width $N_w$. The black solid line with circles and the red squares denote the numerical calculation from the tight-binding Hamiltonian with all the hoppings $t_2$ and the analytical result from $E_g=\frac{9(1+2p)t^2_2}{2|t_1|(N_w+1)^2}$, respectively. Here, $t_1=-2.88\ $eV, $t_2=0.22\ $eV and $t_3=\Delta t_1=U=0$.}
\label{fig: energy gap t2}
\end{figure}

 For $N_w>2$, we show in Fig.~\ref{fig: unit cell t2 gamma point}(a) that the Hamiltonian of the unit cell of the $N_w$-atom wide armchair nanoribbon with the hoppings $t_2$ $H^{\Gamma}(t_2)$ is equivalent to direct product of a Hamiltonian of an $N_w$-site 1D chain $H^{\Gamma}_{\rm 1D}(t_2)$ and an identity matrix $\sigma_0$, i.e., $H^{\Gamma}(t_2)=H^{\Gamma}_{\rm 1D}(t_2)\bigotimes\sigma_0$. The $N_w$-site 1D chain $H^{\Gamma}_{\rm 1D}(t_2)$ only includes the second NN hopping $t_2$ labelled by the curved double-headed arrows, with more details in Sec.~\ref{sect:A1} of the Supplementary Information. For two zero-energy states, we have $H^{\Gamma}(t_2)\Phi_{\pm,2p+2}=H^{\Gamma}_{\rm 1D}(t_2)\Phi_{2p+2}\bigotimes \sigma_0\ket{\pm}$. As shown in Fig.~\ref{fig: unit cell t2 gamma point}(b), similar to the case of the hopping $t_3$, $H^{\Gamma}_{\rm 1D}(t_2)\Phi_{2p+2}$ also decouples into $p+2$ parts, i.e., two parts at the edges with zero on-site energy, and $p$ equivalent parts in the bulk region with $t_2\eta_x\Big(\phi_{2p+2}(x_{3j-1}),\phi_{2p+2}(x_{3j+1})\Big)^T=t_2\eta_x C_{N_w}(-1,1)^T=-t_2\Big(\phi_{2p+2}(x_{3j-1}),\phi_{2p+2}(x_{3j+1})\Big)^T$. To simplify the calculation, we artificially split zero on-site energy at the edges to one with $-t_2$ indicated by the gray circles and another with $t_2$ indicated by the black circles. The on-site energy $-t_2$ at the edges together with the the hopping $t_2$ in bulk region $t_2\eta_x$ give rise to an overall energy shift of $-t_2$ to the state $\Phi_{2p+2}$ and therefore $E_{\pm,2p+2}(t_2)=\Phi^{\dagger}_{\pm,2p+2}H^{\Gamma}(t_2)\Phi_{\pm,2p+2}=\Phi^{\dagger}_{2p+2}H^{\Gamma}_{\rm 1D}(t_2)\Phi_{2p+2}=-t_2$. For the on-site energy $t_2$ at the edges, in the first order of $t_2$, we have $E^{(1)}_{\pm,2p+2}(t_2)=\phi^{\dagger}_{2p+2}(x_1)t_2\phi_{2p+2}(x_1)+\phi^{\dagger}_{2p+2}(x_{N_w})t_2\phi_{2p+2}(x_{N_w})=\frac{3t_2}{N_w+1}$, which also gives rise to an overall energy shift for two zero-energy states. In the second order of $t_2$, the contribution of the on-site energy $t_2$ at the edges to two zero-energy states is given by
\begin{equation}
E^{(2)}_{\pm,2p+2}(t_2)=2\sum_{\substack{1\le m\le N_w \\ m\neq 2p+2}} \frac{\phi^{\dagger}_{2p+2}(x_1)t_2\phi_{m}(x_1)\phi^{\dagger}_{m}(x_1)t_2\phi_{2p+2}(x_1)}{-E_{\pm,m}}
=\mp\frac{9(1+2p)t^2_2}{4t_1(N_w+1)^2}.
\end{equation}
The prefactor of 2 arises from the contribution of two edges. Therefore, the hopping $t_2$ term not only contributes to an overall energy shift of $-t_2+\frac{3t_2}{N_w+1}$ but also an energy gap opening of $E_g(t_2)=|E^{(2)}_{+,2p+2}(t_2)-E^{(2)}_{-,2p+2}(t_2)|=\frac{9(1+2p)t^2_2}{2|t_1|(N_w+1)^2}$. This corrects the previous understanding that the hopping $t_2$ is inactive to open a gap~\cite{PhysRevLett.130.026401}.  We verify both the overall energy shift $-3t_2+\frac{3t_2}{N_w+1}$ and the gap opening $E_g(t_2)$ for $N_w>2$ with direct numerical calculations as shown in Fig.~\ref{fig: energy gap t2}. The analytical and numerical results almost coincide for the energy shift and remain in close agreement for the gap opening at large $N_w$.

\subsection*{Parity of wavefunctions at the \texorpdfstring{$\Gamma$}{Gamma} and \texorpdfstring{$X$}{X} points}
At the $\Gamma$ point, under mirror symmetry $\mathcal{M}_y:\ (x,y)\mapsto(x,-y)$ with $y$ the periodic direction, the states $\Phi_{+,m}$ $(\Phi_{-,m})$ have even (odd) parity since $\Phi_m$ remain invariant and $\ket{\pm}$ becomes $\pm\ket{\pm}$. Then, the number of the occupied states with odd parity is determined by $E_{-,m}<0$. One obtains $\prod_nP_n(\Gamma)=(-1)^{\lfloor \frac{N_w}{3} \rfloor}$ for $t^e_3+\Delta t_1<0$. Under inversion symmetry $\mathcal{I}:\ (x,y)\mapsto(-x,-y)$, $\Phi_m$ becomes $(-1)^{m+1}\Phi_m$ and $\ket{\pm}$ becomes $\pm\ket{\pm}$, which indicates that the states $\Phi_{+,m}$ with even $m$ and $\Phi_{-,m}$ with odd $m$ have odd parity. As a result, the number of the occupied states with odd parity is determined by $E_{+,m}<0$ with even $m$ and $E_{-,m}<0$ with odd $m$, leading to $\prod_nP_n(\Gamma)=(-1)^{\lfloor \frac{N_w-1}{2} \rfloor+1}$ for $t^e_3+\Delta t_1<0$. As $t^e_3+\Delta t_1$ changes from negative to positive via edge dynamics, the two near-zero-energy states of opposite parity undergo a band inversion. Consequently, the parity product $\prod_nP_n(\Gamma)$ reverses sign across the transition.

\begin{figure}[htbp]
\includegraphics[width=1.0\columnwidth]{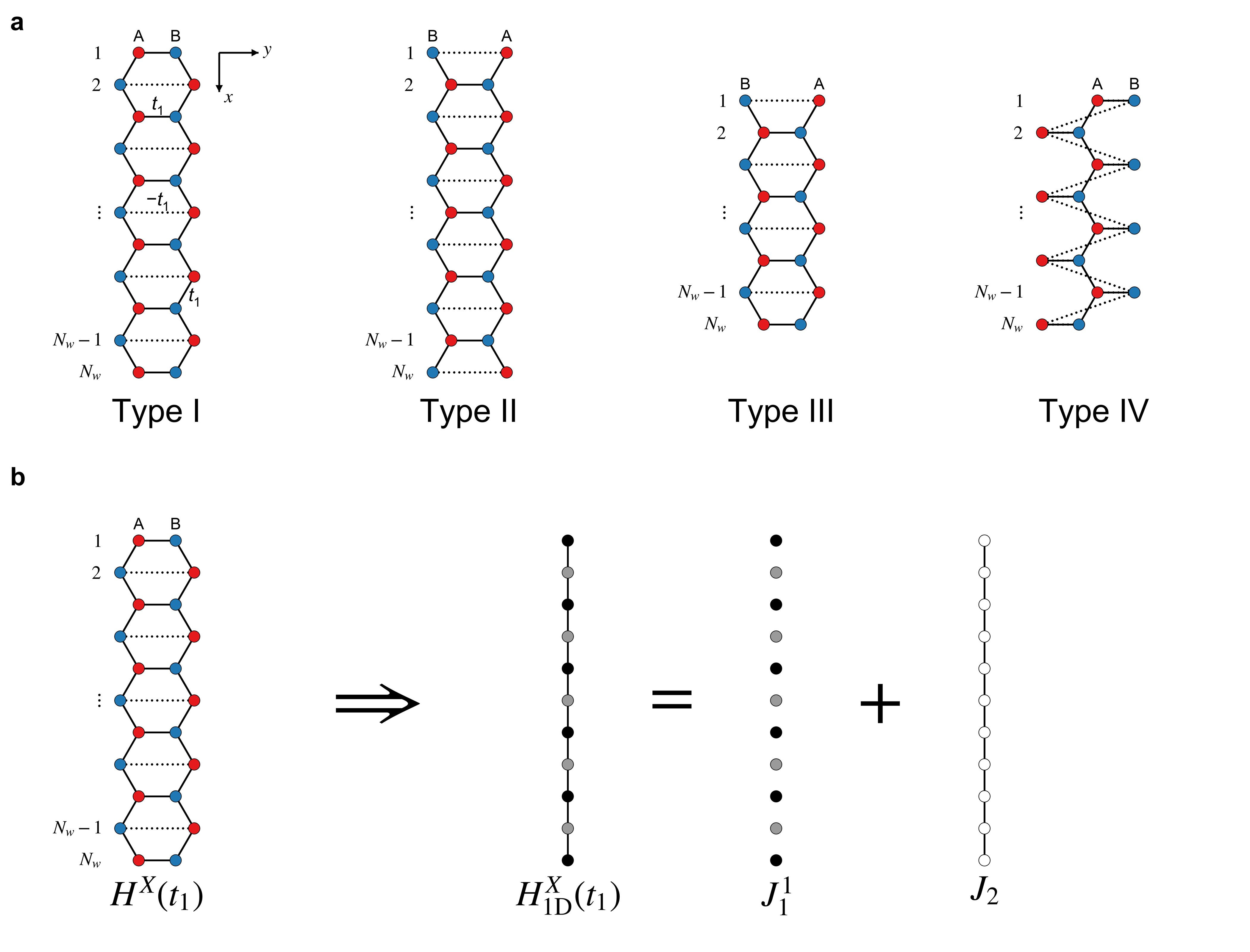}
\caption{(a) AGNR unit cell of four termination types with the first NN hopping at $X$ point. The Red and blue circles represent the A and B sublattices, respectively and the solid (dotted) lines denote the hopping amplitude $t_1$ ($-t_1$). Types I-III possess left-right mirror symmetry, $\mathcal{M}_y:\ (x,y)\mapsto(x,-y)$, whereas Type I, II and IV possess inversion symmetry, $\mathcal{I}:\ (x,y)\mapsto(-x,-y)$. Note that unit cell of Type I and II (III and IV) has odd (even) number of ribbon width $N_w$. (b) Mapping of Hamiltonian of Type-I unit cell with the first NN hopping at $X$ point $H^{X}(t_1)$ to an $N_w$-site 1D Hamiltonian $H^{X}_{\rm 1D}(t_1)=J^1_1+J_2$. $J^1_1$ represents the diagonal contributions with the black (gray) circles on-site energy $t_1$ $(-t_1)$ whereas $J_2$ denotes the off-diagonal first NN hopping $t_1$ shown by the solid lines. The open circles in $J_2$ indicate zero on-site energy.}
\label{fig: unit cell X point}
\end{figure}

At the $X$ point, we only take into account the first NN hopping $t_1$ since the changes of $t^e_3+\Delta t_1$ are not large enough to induce the band inversion. For mirror symmetry, there are three different termination-type induced unit cells as shown in Fig.~\ref{fig: unit cell X point}(a). Both Type I and II have odd $N_w$ whereas Type III has even $N_w$. The Hamiltonian of all these three unit cells with the hoppings $t_1$ $H^X(t_1)$ is equivalent to direct product of a Hamiltonian of a 1D chain $H_{\rm 1D}^X(t_1)$ and a Pauli matrix $\sigma_x$ with $H^X(t_1)=H_{\rm 1D}^X(t_1)\bigotimes\sigma_x$. The mapping of the Hamiltonian of Type I unit cell is shown in Fig.~\ref{fig: unit cell X point}(b) for example. The Hamiltonian of the 1D chain can be written as
\begin{eqnarray}
H^X_{\rm 1D}(t_1)=t_1\left(\begin{array}{ccccc}
1 & 1 & 0      & \cdots & 0      \\
1 & -1 & 1 & \ddots & \vdots \\
0      & 1 & 1 & \ddots & 0      \\
\vdots & \ddots & \ddots & \ddots & 1 \\
0      & \cdots & 0      & 1 & 1
\end{array}\right).\label{eqn: t1 X point}
\end{eqnarray}
Here, the 1D chain includes the first NN hopping $t_1$ and the on-site energy $t_1$ or $-t_1$. We split $H^X_{\rm 1D}(t_1)=J_1+J_2$ with the matrix $J_1$ and $J_2$ representing the Hamiltonian with the on-site energy and the hoppings, respectively. $J_2$ remains the same configuration for three different types of unit cell whereas for Type I, $J^1_1=t_1\,\mathrm{diag}(1,-1,1,-1,\ldots,1,-1,1)$; for Type II, $J^2_1=t_1\,\mathrm{diag}(-1,1,-1,1,\ldots,-1,1,-1)$ and for Type III, $J^3_1=t_1\,\mathrm{diag}(1,-1,1,-1,\ldots,1,-1)$.
Since $[J_2, H_{\rm 1D}^{\Gamma}(t_1)]=0$, the eigenvalues and the eigenstates of $J_2$ are 
$\tilde{E}_m=2t_1\cos(\frac{\pi m}{N_w+1})$ and $\Phi_{m}$, respectively. Further, due to  $\{J^i_1, J_2\}=0$, 
we have $J^i_1\Phi_{m}=(-1)^{i+1}t_1\Phi_{N_w+1-m}$ ($i=1,2,3$) where $\tilde{E}_m=-\tilde{E}_{N_w+1-m}$ and therefore $J^i_1$ couples $\Phi_{m}$ and $\Phi_{N_w+1-m}$ with 
$\Phi^{\dagger}_{m}J^i_1\Phi_{N_w+1-m}=(-1)^{i+1}t_1$. For the unit cell of Type III, due to even $N_w$, $\tilde{E}_m$ is always nonzero and therefore the states $\Phi_{m}$ and $\Phi_{N_w+1-m}$ always come in pairs. This leads to the resulting eigenvalues and eigenstates of $H^X_{\rm 1D}(t_1)$ with $\tilde{E}_{\beta,m}=\beta\lambda_m$ and $\tilde{\Phi}_{\beta,m}=\beta\sqrt{\frac{\lambda_m+\beta\tilde{E}_m}{2\lambda_m}}\Phi_m+\sqrt{\frac{\lambda_m-\beta\tilde{E}_m}{2\lambda_m}}\Phi_{N_w+1-m}$ ($1\le m \le \frac{N_w}{2}$). Here, $\beta=\pm$ and $\lambda_m=\sqrt{\tilde{E}_m^2+t_1^2}$. As a result, the eigenvalues and eigenstates of $H^X(t_1)$ are $\tilde{E}_{\alpha,\beta,m}=\alpha\beta\lambda_m$ and $\tilde{\Phi}_{\alpha,\beta,m}=\tilde{\Phi}_{\beta,m}\bigotimes \ket{\alpha}$ with $\alpha=\pm$. It is evident that these eigenstates are always doubly degenerate with one even parity ($\alpha=+$) and another odd parity ($\alpha=-$). Therefore, the number of the occupied states with odd parity is $\frac{N_w}{2}$, leading to $\prod_nP_n(X)=(-1)^{\frac{N_w}{2}}$. Similarly, for the unit cell of Type I and II, there are $N_w-1$ pairs of doubly degenerate eigenstates with one even parity and another odd parity. Half of them are occupied, giving rise to $\frac{N_w-1}{2}$ occupied states with odd parity. In addition, when $m=\frac{N_w+1}{2}$ and $\tilde{E}_{\frac{N_w+1}{2}}=0$, the remaining unpaired $\Phi_{\frac{N_w+1}{2}}$ is the eigenstate of $J^i_1$ and also $H^X_{\rm 1D}(t_1)$ with the eigenvalue $(-1)^{i+1}t_1$. For $H^X(t_1)$, the corresponding eigenvalue and eigenstate are $\tilde{E}_{\alpha,\frac{N_w+1}{2}}=\alpha(-1)^{i+1}t_1$ and $\tilde{\Phi}_{\alpha,\frac{N_w+1}{2}}={\Phi}_{\frac{N_w+1}{2}}\bigotimes \ket{\alpha}$, respectively. The number of occupied state with odd parity is determined by $\tilde{E}_{-,\frac{N_w+1}{2}}<0$, which is only valid when $i=2$ for Type II unit cell. Therefore, we have $\prod_nP_n(X)=(-1)^{\frac{N_w-1}{2}}$ for Type I and $\prod_nP_n(X)=(-1)^{\frac{N_w+1}{2}}$ for Type II.   

For inversion symmetry, there are three different termination-type induced unit cells as shown in Fig.~\ref{fig: unit cell X point}(a). Type I and Type II have odd $N_w$  whereas Type IV has even $N_w$. For the unit cell of Type I, the eigenvalues and eigenstates are shown in the case of mirror symmetry. Under inversion symmetry, $\Phi_m \rightarrow (-1)^{m+1}\Phi_m$ and $\Phi_{N_w+1-m} \rightarrow (-1)^{N_w+1-m+1}\Phi_{N_w+1-m}=(-1)^{m+1}\Phi_{N_w+1-m}$ due to odd $N_w$, which indicates that $\tilde{\Phi}_{\beta,m}\rightarrow (-1)^{m+1}\tilde{\Phi}_{\beta,m}$. Together with $\ket{\alpha}\rightarrow \alpha\ket{\alpha}$, $\tilde{\Phi}_{\alpha,\beta,m}$ have odd parity when $m$ is even and $\alpha=+$ or $m$ is odd and $\alpha=-$. This means that one of the doubly degenerate states has odd parity and therefore we have $\frac{N_w-1}{2}$ occupied states with odd parity. For the unpaired state $\tilde{\Phi}_{\alpha,\frac{N_w+1}{2}}$, it becomes $(-1)^{\frac{N_w+1}{2}+1}\alpha\tilde{\Phi}_{\alpha,\frac{N_w+1}{2}}$, which has odd parity when $N_w \bmod 4 = 1$ and $\alpha=-$ or $N_w \bmod 4 = 3$ and $\alpha=+$. Only when $N_w \bmod 4 = 3$ and $\alpha=+$, the state $\tilde{\Phi}_{+,\frac{N_w+1}{2}}$ is occupied with $\tilde{E}_{+,\frac{N_w+1}{2}}=t_1<0$. Therefore, we have      
\begin{equation}
\prod_nP_n(X)=
\begin{cases}
(-1)^{\frac{N_w-1}{2}}, & N_w \bmod 4 = 1,\\
(-1)^{\frac{N_w+1}{2}}, & N_w \bmod 4 = 3,
\end{cases}
\end{equation}
leading to $\prod_nP_n(X)=1$ for both cases. We apply similar analysis to the unit cell of Type II and obtain 
\begin{equation}
\prod_nP_n(X)=
\begin{cases}
(-1)^{\frac{N_w+1}{2}}, & N_w \bmod 4 = 1,\\
(-1)^{\frac{N_w-1}{2}}, & N_w \bmod 4 = 3.
\end{cases}
\end{equation}
Both give rise to $\prod_nP_n(X)=-1$. For the unit cell of Type IV, things become quite trivial since all the states are doubly degenerate and one of the doubly degenerate states has odd parity. Half of these states with odd parity are occupied and therefore we have $\prod_nP_n(X)=(-1)^{\frac{N_w}{2}}$.  

We list in Table~\ref{tab:topological_invariant} the topological invariant for different unit-cell types associated with mirror symmetry $\mathcal{M}_y$ and inversion symmetry $\mathcal{I}$, for AGNRs belonging to the $N_w=3p+2$ family, in the case of $t^e_3+\Delta t_1<0$. When the sign of $t^e_3+\Delta t_1$ changes, the topological phases are always distinct, therefore giving different $Z_2$ invariant.

\clearpage

\clearpage

\clearpage

\clearpage

\clearpage

\clearpage

\begin{table}[htbp] 
    \centering
    \renewcommand{\arraystretch}{1.5}
    \caption{Topological invariant $(-1)^{Z_2}$ for different unit-cell
    types associated with mirror symmetry $\mathcal{M}_y$ and inversion
    symmetry $\mathcal{I}$, for AGNRs with $N_w=3p+2$, in the case of $t^e_3+\Delta t_1<0$.}
    \label{tab:topological_invariant}
    \begin{tabular}{|c|c|c|c|c|}
        \hline
        \textbf{Symmetry/Unit cell}
        & \textbf{Type I}
        & \textbf{Type II}
        & \textbf{Type III}
        & \textbf{Type IV} \\
        \hline
        $\mathcal{M}_y$
        & $(-1)^{(p+1)/2}$
        & $(-1)^{(p-1)/2}$
        & $(-1)^{p/2+1}$
        & --- \\
        \hline
        $\mathcal{I}$
        & $(-1)^{(p+1)/2}$
        & $(-1)^{(p-1)/2}$
        & ---
        & $+1$ \\
        \hline
    \end{tabular}
\end{table}

\clearpage
\begin{figure}[htbp]
\includegraphics[width=1.0\columnwidth]{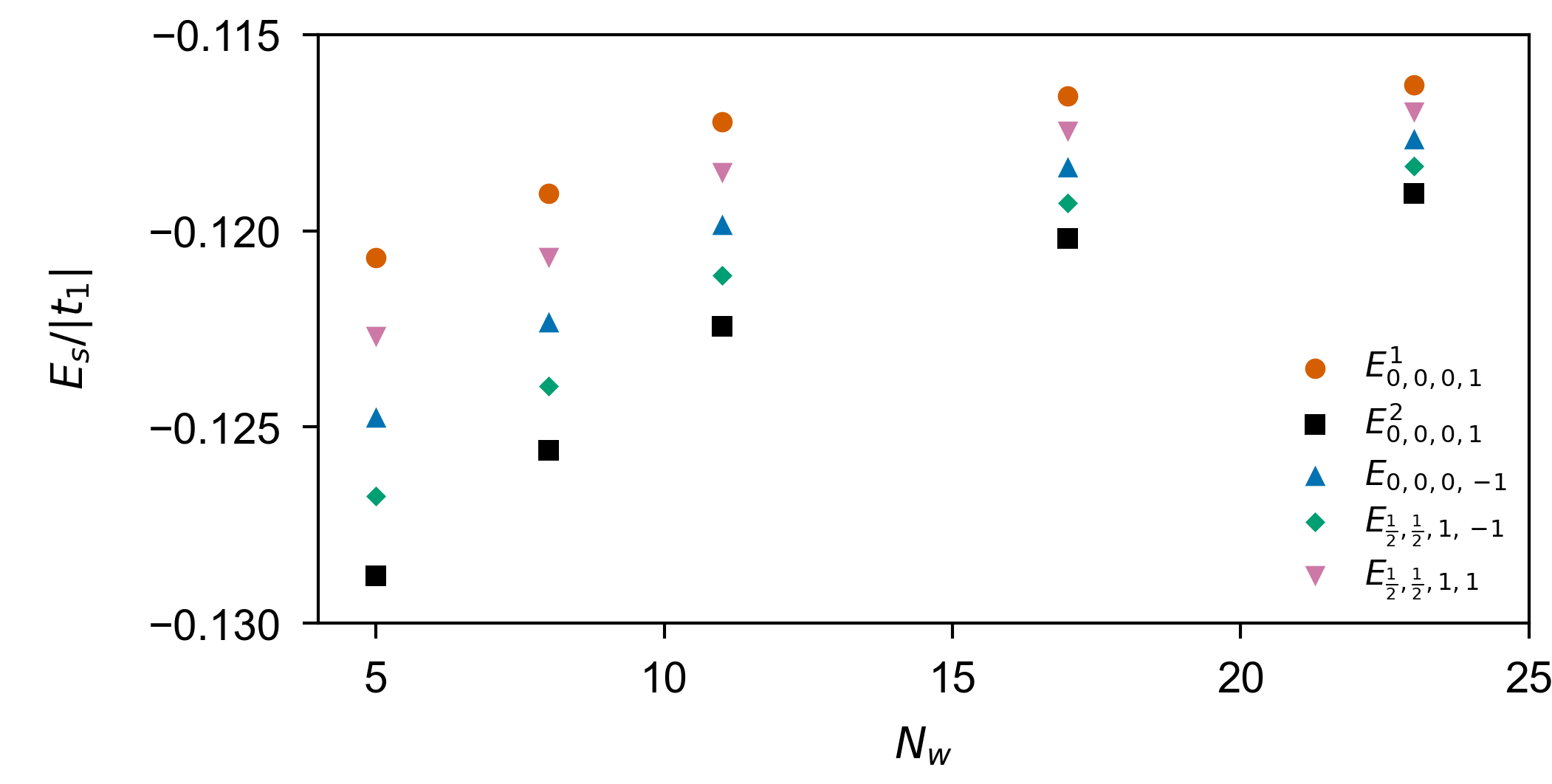}
\caption{Energy shift $E_s$ due to second order effect of Hubbard interactions as a function of armchair ribbon width $N_w$. For $Q=0$, the energies of three singlets $E^1_{0,0,0,1}$, $E^2_{0,0,0,1}$ and $E_{0,0,0,-1}$ are shown. The three triplets are not presented since they have the same energy as the singlet $E_{0,0,0,-1}$. For $Q=1$, the energies of two states with total spin $S=\frac{1}{2}$ and spin projection $S_z=\frac{1}{2}$ $E_{\frac{1}{2},\frac{1}{2},1,\pm 1}$ are shown. Here, $t_1=-2.88\ $eV, $U/|t_1|=1.0$ and $t_2=t_3=\Delta t_1=0$.}
\label{fig: interaction}
\end{figure}

\clearpage
\section*{Supplementary information}

\subsection{Hamiltonian of \texorpdfstring{$N_w$}{Nw}-site 1D chain\label{sect:A1}}
The detailed Hamiltonian of $N_w$-site 1D chain $H^{\Gamma}_{\rm 1D}(t_1)$, $H^{\Gamma}_{\rm 1D}(t_3)$, $H^{\Gamma}_{\rm 1D}(2t_2)$, and $H^{\Gamma}_{\rm 1D}(t_2)$ are written as 
\begin{eqnarray}
H^{\Gamma}_{\rm 1D}(t_1)=t_1\left(\begin{array}{ccccc}
1 & 1 & 0      & \cdots & 0      \\
1 & 1 & 1 & \ddots & \vdots \\
0      & 1 & 1 & \ddots & 0      \\
\vdots & \ddots & \ddots & \ddots & 1 \\
0      & \cdots & 0      & 1 & 1
\end{array}\right),\label{eqn: t1}
\end{eqnarray}

\begin{eqnarray}
H^{\Gamma}_{\rm 1D}(t_3)=t_3\left(\begin{array}{cccccc}
1 & 0 & 1    & 0 & \cdots & 0      \\
0 & 1 & 0 & 1 & \ddots & \vdots \\
1      & 0 & 1 & \ddots & \ddots & 0      \\
0 & 1 & 0 & \ddots & 0 & 1                   \\
\vdots & \ddots & \ddots & \ddots & 1 & 0 \\
0      & \cdots & 0      & 1 & 0 & 1
\end{array}\right),\label{eqn: t3}
\end{eqnarray}

\begin{eqnarray}
H^{\Gamma}_{\rm 1D}(2t_2)=2t_2\left(\begin{array}{ccccc}
0 & 1 & 0      & \cdots & 0      \\
1 & 0 & 1 & \ddots & \vdots \\
0      & 1 & 0 & \ddots & 0      \\
\vdots & \ddots & \ddots & \ddots & 1 \\
0      & \cdots & 0      & 1 & 0
\end{array}\right),\label{eqn: 2t2}
\end{eqnarray}

and 
\begin{eqnarray}
H^{\Gamma}_{\rm 1D}(t_2)
=t_2
\left(
\begin{array}{cccccc}
0 & 0 & 1 & 0 & \cdots & 0 \\
0 & 0 & 0 & 1 & \ddots & \vdots \\
1 & 0 & 0 & \ddots & \ddots & 0 \\
0 & 1 & 0 & \ddots & 0 & 1 \\
\vdots & \ddots & \ddots & \ddots & 0 & 0 \\
0 & \cdots & 0 & 1 & 0 & 0
\end{array}
\right),
\label{eqn:t2}
\end{eqnarray}
respectively.

\subsection{Energy gap opening due to Hubbard interaction\label{sect:A3}}
As introduced in Eq.~\eqref{eqn:hubbard H}, 
$H =\sum_{\langle{i,j}\rangle,s}t_1 c^\dagger_{i,s} c_{j,s}-\frac{U}{2}\sum_i(n_{i,\uparrow}-n_{i,\downarrow})^2$ where only the first NN hopping is considered for simplicity. To solve this many-body Hamiltonian, we first diagonalize the non-interacting part 
by performing a unitary transformation and then obtain $H_0=\sum_{k,s}\epsilon_{k}c^{\dagger}_{k,s}c^{}_{k,s}$ with $\epsilon_{k}$ the eigenvalues, $c^{}_{k,s}=\sum_{x}O_{k;x}c^{}_{x,s}$ and $k$ the index of different eigenstates. The unitary matrix $O$ can be constructed from the eigenfunctions.  

%After expanding the quadratic term, this Hamiltonian becomes $H =\sum_{\langle{i,j}\rangle,s}t_1 c^\dagger_{i,s} c_{j,s}-\frac{U}{2}\sum_i\left(c^\dagger_{i,\uparrow}c^{}_{i,\uparrow}+c^\dagger_{i,\downarrow}c^{}_{i,\downarrow}\right)+U\sum_i c^\dagger_{i,\downarrow}c^\dagger_{i,\uparrow}c^{}_{i,\uparrow}c^{}_{i,\downarrow}\equiv H_0+H_1+H_2$.

The non-interacting half-filled state is defined by 
\begin{equation}
|\Omega\rangle=\left(\prod_{k\ s.t.\ \epsilon_k\leq 0,s}c^\dagger_{k,s}\right)|0\rangle
\end{equation}
where $|0\rangle$ is the vacuum state with the definition such that $c_{k,s}|0\rangle=0\ \forall\ k,s$. Since there are four states hosting zero energy, we now have six many-body non-interacting degenerate states at half filling. Without loss of generality, we assume two zero-energy states with odd parity under mirror symmetry are occupied in the state $|\Omega\rangle$. For convenience, we further introduce the following operators 
\begin{equation}
c_{k,s}=
\begin{cases}
a^{}_{k,s} & \forall\ k\ s.t.\ \epsilon_k\geq 0 \quad{\rm ``particles"}\\
b^{\dagger}_{-k,s}& \forall\ k\ s.t.\ \epsilon_k\equiv\epsilon_{-k}\leq 0 \quad {\rm ``holes"}\ .
\end{cases}
\end{equation}
With this definition it is clear that 
\begin{equation}
a^{}_{k,s}|\Omega\rangle=b^{}_{-k,s}|\Omega\rangle=0\ \forall\ k,s.
\end{equation}
Specially, $c^\dagger_{d,s,+}|0\rangle \equiv a^\dagger_{d,s}|0\rangle$ and $c^\dagger_{d,s,-}|0\rangle\equiv b^{}_{-d,s}|0\rangle$ where $d$ denotes 
the zero-energy degenerate state, $s=\uparrow,\downarrow$, and the $\pm$ index denotes the parity of the state under mirror symmetry. With such operators of particles and holes, one can rewrite the above Hamiltonian as 
\begin{align}
%H&=H_0+H_1+H_2\\
H&=\epsilon_{k}a^{\dagger}_{k,s}a^{}_{k,s}-\epsilon_{-k}b^{\dagger}_{-k,s}b^{}_{-k,s}+\mathcal{E}_{\Omega}-\frac{U}{2}\Lambda\\
&+V_{k',l';k,l}a^\dagger_{k',\uparrow}a^\dagger_{l',\downarrow}a^{}_{k,\downarrow}a^{}_{l,\uparrow}-V_{k',l';k,-l}a^\dagger_{k',\uparrow}a^\dagger_{l',\downarrow}b^\dagger_{-l,\uparrow}a^{}_{k,\downarrow}\\
&+V_{k',l';-k,l}a^\dagger_{k',\uparrow}a^\dagger_{l',\downarrow}b^\dagger_{-k,\downarrow}a^{}_{l,\uparrow}+V_{k',-l';k,l}a^\dagger_{k',\uparrow}b^{}_{-l',\downarrow}a^{}_{k,\downarrow}a^{}_{l,\uparrow}\\
&-V_{-k',l';k,l}a^\dagger_{l',\downarrow}b^{}_{-k',\uparrow}a^{}_{k,\downarrow}a^{}_{l,\uparrow}+V_{k',l';-k,-l}a^\dagger_{k',\uparrow}a^\dagger_{l',\downarrow}b^\dagger_{-k,\downarrow}b^\dagger_{-l,\uparrow}\\
&+V_{k',-l';k,-l}a^\dagger_{k',\uparrow}b^\dagger_{-l,\uparrow}b^{}_{-l',\downarrow}a^{}_{k,\downarrow}-V_{-k',l';k,-l}a^\dagger_{l',\downarrow}b^\dagger_{-l,\uparrow}b^{}_{-k',\uparrow}a^{}_{k,\downarrow}\\
&-V_{k',-l';-k,l}a^\dagger_{k',\uparrow}b^\dagger_{-k,\downarrow}b^{}_{-l',\downarrow}a^{}_{l,\uparrow}+V_{-k',l';-k,l}a^\dagger_{l',\downarrow}b^\dagger_{-k,\downarrow}b^{}_{-k',\uparrow}a^{}_{l,\uparrow}\\
&+V_{-k',-l';k,l}b^{}_{-k',\uparrow}b^{}_{-l',\downarrow}a^{}_{k,\downarrow}a^{}_{l,\uparrow}+V_{k',-l';-k,-l}a^\dagger_{k',\uparrow}b^\dagger_{-k,\downarrow}b^\dagger_{-l,\uparrow}b^{}_{-l',\downarrow}\\
&-V_{-k',l';-k,-l}a^\dagger_{l',\downarrow}b^\dagger_{-k,\downarrow}b^\dagger_{-l,\uparrow}b^{}_{-k',\uparrow}-V_{-k',-l';k,-l}b^\dagger_{-l,\uparrow}b^{}_{-k',\uparrow}b^{}_{-l',\downarrow}a^{}_{k,\downarrow}\\
&+V_{-k',-l';-k,l}b^\dagger_{-k,\downarrow}b^{}_{-k',\uparrow}b^{}_{-l',\downarrow}a^{}_{l,\uparrow}+V_{-k',-l';-k,-l}b^\dagger_{-k,\downarrow}b^\dagger_{-l,\uparrow}b^{}_{-k',\uparrow}b^{}_{-l',\downarrow}\\
&\equiv \tilde H_0+\tilde V
\end{align}
where $\tilde H_0=\epsilon_{k}a^{\dagger}_{k,s}a^{}_{k,s}-\epsilon_{-k}b^{\dagger}_{-k,s}b^{}_{-k,s}+\mathcal{E}_{\Omega}-\frac{U}{2}\Lambda$ and $\tilde V$ includes all terms with $V$. Here, $\mathcal{E}_{\Omega}=\sum_{k,s}\epsilon_{-k}=2\sum_{k}\epsilon_{-k}$, $\Lambda=N_x/2$ with $N_x$ the number of lattice points and $V_{k',l';k,l}=U\sum_xO^{}_{k',x}O^{}_{l',x}O^{-1}_{x,k}O^{-1}_{x,l}$. In the case of weak interactions, all $\tilde V$ terms can be treated as perturbations.

\begin{table}[h]
\centering
\renewcommand{\arraystretch}{1.3}
\begin{tabular}{c c c c c}
\hline
States $|\Omega_{S,S_z,Q, \Pi}\rangle$ & $S$ & $S_z$ & $Q$ & $\Pi$ \\
\hline

$b^\dagger_{-d,\uparrow}b^\dagger_{-d,\downarrow}|\Omega\rangle$
& 0 & 0 & -2 & +1 \\

$b^\dagger_{-d,s}|\Omega\rangle$
& $\frac{1}{2}$ & $\pm \frac{1}{2}$ & -1 & -1 \\

$a^\dagger_{d,s}b^\dagger_{-d,\uparrow}b^\dagger_{-d,\downarrow}|\Omega\rangle$
& $\frac{1}{2}$ & $\pm \frac{1}{2}$ & -1 & +1 \\

$|\Omega\rangle \;,\; a^\dagger_{d,\uparrow}a^\dagger_{d,\downarrow}b^\dagger_{-d,\uparrow}b^\dagger_{-d,\downarrow}|\Omega\rangle$
& 0 & 0 & 0 & +1 \\

$\frac{1}{\sqrt{2}}\left(a^\dagger_{d,\uparrow}b^\dagger_{-d,\uparrow}+a^\dagger_{d,\downarrow}b^\dagger_{-d,\downarrow}\right)|\Omega\rangle$
& 0 & 0 & 0 & -1 \\

$\frac{1}{\sqrt{2}}\left(a^\dagger_{d,\uparrow}b^\dagger_{-d,\uparrow}-a^\dagger_{d,\downarrow}b^\dagger_{-d,\downarrow}\right)|\Omega\rangle$
& 1 & 0 & 0 & -1 \\

$a^\dagger_{d,\uparrow}b^\dagger_{-d,\downarrow}|\Omega\rangle \;\; a^\dagger_{d,\downarrow}b^\dagger_{-d,\uparrow}|\Omega\rangle$
& 1 & $\pm 1$ & 0 & -1 \\

$b^\dagger_{-d,s}a^\dagger_{d,\uparrow}a^\dagger_{d,\downarrow}|\Omega\rangle$
& $\frac{1}{2}$ & $\pm \frac{1}{2}$ & +1 & -1 \\

$a^\dagger_{d,s}|\Omega\rangle$
& $\frac{1}{2}$ & $\pm \frac{1}{2}$ & +1 & +1 \\

$a^\dagger_{d,\uparrow}a^\dagger_{d,\downarrow}|\Omega\rangle$
& 0 & 0 & +2 & +1 \\
\hline
\end{tabular}
\caption{Many-body non-interacting states $|\Omega_{S,S_z,Q, \Pi}\rangle$ constructed from $a^\dagger_{d,s}$ and $b^\dagger_{-d,s}$ operators acting on $|\Omega\rangle$. Note that we define $|\Omega\rangle\equiv|\Omega^1_{0,0,0,1}\rangle$ and $a^\dagger_{d,\uparrow}a^\dagger_{d,\downarrow}b^\dagger_{-d,\uparrow}b^\dagger_{-d,\downarrow}|\Omega\rangle\equiv|\Omega^2_{0,0,0,1}\rangle$ since these two states are indistinguishable using the quantum numbers $S$, $S_z$, $Q$ and $\Pi$.}
\label{tab:quantum_numbers}
\end{table}

The half-filled state $|\Omega\rangle$ becomes 
\begin{equation}
|\Omega\rangle=\left(\prod_{-k,s}b_{-k,s}\right)|0\rangle\quad\forall\ \epsilon_{-k}\le 0\,
\end{equation}
which has 15 many-body non-interacting degenerate states including 5 at half filling mentioned above. All these non-interacting degenerate states with eigenenergy $\mathcal{E}_\Omega-\frac{U}{2}\Lambda$ are listed explicitly in Table~\ref{tab:quantum_numbers} labeling by different total spin $S$, spin projection $S_z$, charge $Q$, and parity $\Pi$. The interaction $\tilde V$ does not violate these quantum numbers. To see how the interactions affect the energy shift of these many-body states, we perform degenerate perturbation theory on the states with $Q=0$ and $Q=1$ up to the second order of the interactions. For the first order, the contributions of $\tilde V$ are always zero, which agree with the previous calculations. We then turn to calculate the second order using 
\begin{equation}
\sum_{|n\rangle\ne |\Omega_{S,S_z,Q, \Pi}\rangle} \langle\Omega_{S,S_z,Q, \Pi}|\tilde V|n\rangle\frac{1}{\langle\Omega_{S,S_z,Q, \Pi}|\tilde H_0|\Omega_{S,S_z,Q, \Pi}\rangle-\langle n|\tilde H_0|n\rangle}\langle n|\tilde V|\Omega_{S,S_z,Q, \Pi}\rangle.
\end{equation}

For $Q=0$, i.e., half-filled states, there are three singlet states $|\Omega^1_{0,0,0,1}\rangle$, $|\Omega^2_{0,0,0,1}\rangle$ and $|\Omega_{0,0,0,-1}\rangle$ as well as three triplet states $|\Omega_{1,0,0,-1}\rangle$ and $|\Omega_{1,\pm 1,0,-1}\rangle$. The corresponding energy shift due to the second order effect is 
\begin{eqnarray}
E^1_{0,0,0,1}&=&-4\sum_{k,l,m\ne d}\frac{V_{k,l;m,d}V_{d,m;l,k}}{\epsilon_{l}+\epsilon_{k}+\epsilon_m}, \\
E^2_{0,0,0,1}&=&-4\sum_{k,l,m\ne d}\frac{V_{k,l;m,-d}V_{-d,m;l,k}}{\epsilon_{l}+\epsilon_{k}+\epsilon_m}, \\
E_{0,0,0,-1}&=&-2\sum_{k,l,m\ne d}\frac{V_{k,l;m,d}V_{d,m;l,k}}{\epsilon_{l}+\epsilon_{k}+\epsilon_m}-2\sum_{k,l,m\ne d}\frac{V_{k,l;m,-d}V_{-d,m;l,k}}{\epsilon_{l}+\epsilon_{k}+\epsilon_m}, \\
E_{1,0,0,-1}=E_{1,\pm 1,0,-1}&=&-2\sum_{k,l,m\ne d}\frac{V_{k,l;m,d}V_{d,m;l,k}}{\epsilon_{l}+\epsilon_{k}+\epsilon_m}-2\sum_{k,l,m\ne d}\frac{V_{k,l;m,-d}V_{-d,m;l,k}}{\epsilon_{l}+\epsilon_{k}+\epsilon_m},
\end{eqnarray}
where none of the indices can be equal to $d$ and all three triplet states share the same energy due to SU(2) symmetry. Note that we ignore the term $-\sum_{k,l,m,p\ne d}\frac{V_{k,l;m,p}V_{p,m;l,k}}{\epsilon_{k}+\epsilon_{l}+\epsilon_m+\epsilon_p}$ throughout the paper since it is universal for all the states. With these energies, we can identify the half-filled ground state. 

For $Q=1$, i.e., single-particle excited states, there are four states with total spin $S=\frac{1}{2}$ and we choose two of them with $S_z=\frac{1}{2}$, i.e., 
$|\Omega_{\frac{1}{2},\frac{1}{2},1,1}\rangle$ and $|\Omega_{\frac{1}{2},\frac{1}{2},1,-1}\rangle$ by considering the SU(2) symmetry. The corresponding energy shift due to the second order effect is given by 
\begin{eqnarray}
E_{\frac{1}{2},\frac{1}{2},1,1}&=&-3\sum_{k,l,m\ne d}\frac{V_{k,l;m,d}V_{d,m;l,k}}{\epsilon_{l}+\epsilon_{k}+\epsilon_m}
-\sum_{k,l,m\ne d}\frac{V_{k,l;m,-d}V_{-d,m;l,k}}{\epsilon_{l}+\epsilon_{k}+\epsilon_m} \\
E_{\frac{1}{2},\frac{1}{2},1,-1}&=&-\sum_{k,l,m\ne d}\frac{V_{k,l;m,d}V_{d,m;l,k}}{\epsilon_{l}+\epsilon_{k}+\epsilon_m}
-3\sum_{k,l,m\ne d}\frac{V_{k,l;m,-d}V_{-d,m;l,k}}{\epsilon_{l}+\epsilon_{k}+\epsilon_m}.
\end{eqnarray}

%The single-particle excitation gap due to the Hubbard interaction is calculated twice as large as the energy difference between the minimum of these two states and the half-filled ground state.

In Fig.~\ref{fig: interaction} we show the energy shift of many-body states with the charge $Q=0$ and $Q=1$ due to second order effect of on-site Hubbard interaction as a function of ribbon width. For half-filled states with $Q=0$, the singlet $|\Omega^2_{0,0,0,1}\rangle$ always has lowest energy and therefore is the ground state. Note that the energies of three triplets with $Q=0$ are not shown since they are exactly the same as the singlet $|\Omega_{0,0,0,-1}\rangle$. For single excited states with $Q=1$, $|\Omega_{\frac{1}{2},\frac{1}{2},1,-1}\rangle$ has lower energy and therefore the single-particle excitation gap is determined by $E_g(U)=2(E_{\frac{1}{2},\frac{1}{2},1,-1}-E^2_{0,0,0,1})$ as shown in Fig.~\ref{fig: energy gap}. We find that the gap due to the Hubbard interaction is at least an order of magnitude smaller than the one induced by the hopping $t^e_3+\Delta t_1$ at the edges.

\end{document}